\documentclass[a4paper,11pt]{article}
\usepackage{jheppub} 
\usepackage{lineno}
\usepackage{amsmath}
\usepackage{graphicx}
\usepackage{subcaption}
\usepackage{multirow}
\usepackage{arydshln}
\usepackage{physics}
\usepackage{bm}
\usepackage{tikz}
\usetikzlibrary{calc}
\usetikzlibrary{fit, positioning}
\usepackage{comment}
\usepackage[colorinlistoftodos]{todonotes}
\usepackage{csquotes}
\usepackage{placeins}
\usepackage{url}
\usepackage[normalem]{ulem}

\newcommand{\mr}{\multirow}
\newcommand{\mc}{\multicolumn}

\newcommand{\nt}{N_\tau}

\newcommand{\ns}{N_\sigma}
\newcommand{\nf}{N_\mathrm{f}}
\newcommand{\amc}{am_\mathrm{c}}
\newcommand{\betac}{\beta_\mathrm{c}}
\newcommand{\betapc}{\beta_\mathrm{pc}}
\newcommand{\ztwo}{\mathrm Z_2}
\newcommand{\pbp}{\bar{\psi}\psi}

\newcommand{\refer}{ref.~}

\newcommand{\eq}{eq.~}

\newcommand{\clqcd}{\texttt{CL\kern-.25em\textsuperscript{2}QCD}}

\usepackage{xcolor}

\title{Testing machine-learned distributions against Monte Carlo data for the QCD chiral phase transition}
\author[a]{Reinhold Kaiser,}
\affiliation[a]{Institut für theoretische Physik,\\
Goethe Universit\"at Frankfurt, D-60323 Frankfurt am Main, Germany}
\author[b]{Frithjof Karsch,}
\affiliation[b]{Fakult\"at f\"ur Physik,\\
Universit\"at Bielefeld, D-33615 Bielefeld, Germany}
\author[a]{Jan Philipp Klinger,}
\author[a]{Owe Philipsen,}
\author[b]{Christian Schmidt,}
\author[c,d]{and Simran Singh}
\affiliation[c]{Transdisciplinary Research Area (TRA) Matter, University of Bonn, Germany}
\affiliation[d]{Helmholtz Institute for Radiation and Nuclear Physics (HISKP), University of Bonn, Germany}

\emailAdd{kaiser@itp.uni-frankfurt.de}
\emailAdd{karsch@physik.uni-bielefeld.de}
\emailAdd{klinger@itp.uni-frankfurt.de}
\emailAdd{philipsen@itp.uni-frankfurt.de}
\emailAdd{schmidt@physik.uni-bielefeld.de}
\emailAdd{ssingh@uni-bonn.de}

\abstract{We demonstrate that conditional Masked Autoregressive Flows constitute a flexible interpolation tool for lattice QCD observables, conditioned on bare lattice parameters. As a benchmark, we use the chiral phase structure of QCD with five degenerate light quark flavours, which on coarse lattices exhibits a region of first-order chiral transitions terminating in a critical quark mass. The method successfully reproduces standard reweighting in the gauge coupling, and naturally extends to interpolation in quark mass and spatial volume, for which reweighting is computationally prohibitive or inapplicable, respectively. Once trained, the model generates samples across the full parameter space in minutes, which can be used to obtain first estimates of the critical quark mass without simulating all intermediate parameter values, consistent with results from reweighted lattice data. This offers a concrete reduction in the number of lattice ensembles required. Precision on the critical mass from learned distributions is so far prohibited by the mode-covering effect inherent to maximum-likelihood-based training, which introduces a systematic bias near first-order transitions. At the current stage, the method is well-suited for a range of practical applications: localising phase boundaries, identifying the universal scaling axes at a critical point, and accelerating informed determinations of parameter values ahead of high-precision Monte Carlo campaigns.}

\begin{document}
\maketitle
\flushbottom

\section{Introduction}\label{sec:intro}
In recent years, considerable effort has been devoted within the lattice QCD community to the application of machine learning (ML) techniques to a variety of challenging problems. These include the generation of gauge configurations using generative models~\cite{Albergo:2019eim, Abbott:2022hkm, Albandea:2023wgd, Abbott:2023thq, Wang:2023exq, Gerdes:2024rjk,Bonanno:2025pdp,Zhu:2025pmw, Lehner:2023prf,Knuttel:2023vvc,Wenger:2026sjp}, the reconstruction of spectral functions~\cite{Zhou:2023pti, Larkoski:2025clo, Wang:2023fhc,Kades:2019wtd,DelDebbio:2024qfh,Buzzicotti:2023qdv}, phase classification~\cite{Wetzel:2017ooo,Bachtis:2020dmf,Bachtis:2020fly,Karsch:2022yka}, as well as more general strategies aimed at accelerating lattice QCD simulations~\cite{Shanahan:2018vcv,Cranmer:2023xbe, Holland:2024muu, Lawrence:2025rnk,Holland:2025fsa, Choi:2026jdv}.
In this work, we consider a further application of ML methods in lattice QCD, namely the interpolation of probability distributions of lattice observables conditioned on the bare parameters of the theory. Particular emphasis is placed on regions of parameter space where standard reweighting techniques~\cite{Ferrenberg:1988yz,Ferrenberg:1989ui} become inefficient or inapplicable. As is common with interpolation approaches of this type, the method used in this work relies on Monte Carlo (MC) time histories generated from lattice simulations at selected parameter points, which serve as the training data.

The data sample for this analysis was generated for \refer\cite{Cuteri:2021ikv}, where the chiral critical surface separating parameter
regions of first-order and crossover transitions was determined.
This surface was mapped in the space spanned by the lattice gauge coupling $\beta$, the number of mass-degenerate
flavours $\nf$, their bare mass $am$, and the lattice spacing parametrised by the temporal lattice extent, $\nt^{-1}=aT$, for 
standard staggered fermions.
The study demonstrates for all $\nf\in[2,6]$ that the first-order chiral phase
transitions, predicted for $\nf\geq 3$ by renormalisation group flows in linear sigma models~\cite{Pisarski:1983ms}, and observed in early numerical studies on coarse lattices~\cite{Brown:1990ev,Iwasaki:1996zt,Karsch:2001nf}, 
weaken with decreasing lattice spacing and disappear before the continuum limit is reached. 
Since then, these studies are getting extended to larger values of $\nf$ up to the onset of the conformal window~\cite{Klinger:2025xxb,Klinger:2026pbe}, as well as to include imaginary chemical potential~\cite{DAmbrosio:2022kig,DAmbrosio:2025ldv}.
Determining the phase diagram as a function of the lattice bare parameters is computationally expensive, involving systematic scans of a high-dimensional parameter space and subsequent finite-size scaling analyses at each parameter set to establish the order of the thermal transition.
This motivates the exploration of techniques to reduce the number of simulations 
and/or the statistics needed to successfully pursue such investigations. 

The work presented here proceeds in this direction and builds upon an earlier study conducted by a subset of the present authors, in which machine learning techniques were employed to analyse the behaviour of thermal transitions in lattice QCD at finite volume and lattice spacing~\cite{Karsch:2022yka,Neumann:2023}. In that study, the first stage consisted of interpolating lattice observables in the gauge coupling~$\beta$, providing an alternative to conventional reweighting methods~\cite{Ferrenberg:1988yz,Ferrenberg:1989ui}. In a subsequent stage, histogram representations of the chiral condensate were used as input to a vision-transformer-based neural network architecture~\cite{dosovitskiy2021imageworth16x16words} to infer the location of a critical phase boundary, in analogy with approaches that extract phase structure from thermodynamic observables such as the equation of state~\cite{Pang:2016vdc}.

In this work, we focus on a systematic validation of the first stage of the previously mentioned ML-based analysis against established numerical methods used in lattice studies. Using existing lattice QCD data generated at multiple parameter sets, we generalise the interpolation from the gauge coupling~$\beta$ to additional directions in parameter space, in particular quark mass $am$ and spatial lattice extent $\ns$. In these additional parameter directions, conventional reweighting techniques~\cite{Ferrenberg:1988yz,Ferrenberg:1989ui} are either computationally demanding or not readily applicable.
As our ML framework, we employ masked autoregressive flows~\cite{Germain:2015yft,Papamakarios:2017tec}, a class of generative models for probability density estimation from samples. The method learns a conditional probability density, with the conditioning variables given by the lattice parameters along which the interpolation is performed. In the language of machine learning, this corresponds to a generative modelling setup, as the model is trained directly on sampled configurations without requiring externally provided target labels.

Clearly, the ML-specific findings of the present work are independent of the particular theory or discretisation employed here,  
and should apply generically to lattice investigations of phase
structures.

\section{Lattice case study: chiral transition with \texorpdfstring{\boldmath$\nf=5$}{Nf=5} degenerate quarks}\label{sec:latt_use_case}

Our choice of sample data set is motivated by the large available
statistics and resulting clean signals, which make it 
a good training and benchmarking set for our ML model.
For moderate $\nt$-values, $\nf=5$ lattice QCD with staggered fermions displays a first-order chiral transition
for small quark masses, 
which weakens and disappears in a $\ztwo$-critical point
as the quark mass increases to its corresponding critical value $\amc$. 
For $am>\amc$ the thermal transition is a crossover. Since there are no non-analytic
phase transitions in finite volume, the histograms for the order parameter look
quite similar for all cases in the vicinity of the critical point, and become 
distinguishable only by an intricate finite-size scaling analysis.
In this section 
we briefly summarise the important aspects of the associated lattice simulations and explain how to determine the critical mass $\amc$ using numerical reweighting techniques and a finite-size scaling formula for the kurtosis of the order parameter distribution.

\subsection{Lattice simulations}

All configurations and measurements used here were already generated for \refer\cite{Cuteri:2021ikv} using the lattice QCD code \clqcd~\cite{sciarra_2021_5121917} for unimproved staggered fermions and the Wilson gauge action, which employs the rational hybrid Monte Carlo (RHMC) algorithm~\cite{Kennedy:1998cu} for degenerate quark flavours.
To reduce the noise of the stochastic estimate of the fermion determinant, the multiple pseudo-fermions technique~\cite{Clark:2006fx} was used.
More detailed information about the Monte Carlo simulations can be found in section 3 of \refer\cite{Cuteri:2021ikv}.

The lattice spacing $a(\beta)$ depends on the inverse gauge coupling $\beta$, which
is used to tune the temperature according to the relation $T=1/(a\nt)$.
Consequently, for a fixed temperature, increasing $\nt$ corresponds to decreasing the lattice spacing.
For a given set of $\nf,\nt,am$, Monte Carlo simulations were first performed for 
two to four $\beta$-values, chosen in the vicinity of and containing the transition point.
For each Monte Carlo trajectory the chiral condensate $\pbp$ was measured as a (quasi-)order parameter for the chiral transition using $16$ stochastic estimators, as well as the plaquette to calculate the gauge action $S_G$.
This allows for a determination of the 
pseudo-critical coupling
$\betapc$, corresponding to the pseudo-critical temperature
at the phase boundary separating the first-order and crossover regions.
This was repeated for several quark masses chosen
around and containing the expected critical mass value $\amc$.
The spatial volume of the lattices is $L^3$ with $L=a\ns$
for all three spatial dimensions.
For each mass, the simulations are then repeated on lattices with different aspect ratios $\ns/\nt\in\{2,3,4\}$.
An overview of the resulting data set and its statistics can be found in 
tables~\ref{tab:nt4-statistics} and~\ref{tab:nt6-statistics}.
More detailed analytics of the simulations can be found in the appendix of \refer\cite{Cuteri:2021ikv}.
\begin{table}[h]
    {
    \setlength{\tabcolsep}{4pt}
    \centering
    \begin{subtable}{0.495\linewidth}
    \centering
    \small{
    \begin{tabular}{|l|ll|ll|ll|}
        \hline
        \mr{2}{*}{$am$}     & \mc{3}{c}{$\beta$}    & \mc{3}{c|}{Total statistics}  \\\cdashline{2-7}
                            & \mc{2}{c|}{$\ns=8$}   & \mc{2}{c|}{$\ns=12$}  & \mc{2}{c|}{$\ns=16$}  \\
        \hline
        \mr{4}{*}{$0.075$} & $4.966$  & 200k      &           &           &           &          \\
                            & $4.968$  & 160k      & $4.968$  & 120k      & $4.968$  & 120k      \\
                            & $4.970$  & 160k      & $4.969$  & 120k      & $4.969$  & 120k      \\
                            & $5.972$  & 120k      & $4.970$  & 119k      & $4.970$  & 120k      \\
        \hline
        \mr{4}{*}{$0.080$} & $4.976$  & 120k      &           &           &           &          \\
                            & $4.978$  & 160k      & $4.978$  & 120k      & $4.978$  & 120k      \\
                            & $4.980$  & 160k      & $4.979$  & 120k      & $4.979$  & 160k      \\
                            & $5.982$  & 120k      & $4.980$  & 120k      & $4.980$  & 120k      \\
        \hline
        \mr{3}{*}{$0.085$} & $4.988$  & 120k      & $4.988$  & 120k      & $4.988$  & 120k       \\
                            & $4.990$  & 120k      & $4.989$  & 120k      & $4.989$  & 120k      \\
                            & $4.992$  & 120k      & $4.990$  & 120k      & $4.990$  & 120k      \\
        \hline
        \mr{4}{*}{$0.090$}  & $4.996$  & 120k      & $4.996$  & 120k      &          &           \\
                            & $4.998$  & 120k      & $4.998$  & 120k      & $4.998$  & 120k      \\
                            & $5.000$  & 120k      & $5.000$  & 120k      & $4.999$  & 120k      \\
                            & $5.002$  & 120k      &          &           &          &           \\
        \hline
    \end{tabular}
    }
    \caption{$\nt=4$}
    \label{tab:nt4-statistics}
    \end{subtable}
    \hfill
    \begin{subtable}{0.495\linewidth}
    \centering
    \small{
    \begin{tabular}{|l|ll|ll|ll|}
        \hline
        \mr{2}{*}{$am$}     & \mc{3}{c}{$\beta$}    & \mc{3}{c|}{Total statistics}  \\\cdashline{2-7}
                            & \mc{2}{c|}{$\ns=12$}   & \mc{2}{c|}{$\ns=18$}  & \mc{2}{c|}{$\ns=24$}  \\
        \hline
        \mr{3}{*}{$0.020$}  & $4.968$  & 160k      & $4.971$  & 120k      &          &           \\
                            & $4.972$  & 120k      & $4.972$  & 159k      &          &           \\
                            & $4.760$  & 120k      & $4.973$  & 160k      &          &           \\
        \hline
        \mr{4}{*}{$0.025$}  &          &           & \mc{2}{c|}{$\ns=16$} &          &          \\\cdashline{4-5}
                            & $4.985$  & 160k      & $4.9875$ & 160k      & $4.987$  & 80k      \\
                            & $4.990$  & 160k      & $4.990$  & 160k      & $4.989$  & 80k      \\
                            & $4.995$  & 160k      & $4.9925$ & 120k      & $4.991$  & 75k      \\
        \hline
        \mr{3}{*}{$0.030$}  & $5.004$  & 120k      & $5.004$  & 120k      & $5.005$  & 100k      \\
                            & $5.006$  & 120k      & $5.006$  & 120k      & $5.006$  & 100k      \\
                            & $5.008$  & 120k      & $5.008$  & 120k      & $5.007$  & 99k       \\
        \hline
    \end{tabular}
    }
    \caption{$\nt=6$}
    \label{tab:nt6-statistics}
    \end{subtable}
    \caption{Simulation statistics for $\nf=5$ data~\cite{Cuteri:2021ikv}. Each column contains the $\beta$-value in the left sub-column and the total statistics in the right sub-column.}\label{tab:Nf5_statistics}
    } 
\end{table}

\subsection[Estimating the \texorpdfstring{$\ztwo$}{Z2}-point]{Estimating the \texorpdfstring{\boldmath$\ztwo$}{Z2}-point}\label{sec:estimating-ztwo}

The method to determine 
phase transitions and their order by simulations on finite volumes 
is standard and based on the finite-size scaling of generalised cumulants
of the order parameter $O=\langle \bar{\psi}\psi\rangle$,
\begin{equation}
	\label{equ:std-moments}
	B_n(\beta; am , \ns) = \frac{\ev{\left( O - \ev{O}\right)^n}}{\ev{\left(O - \ev{O}\right)^2}^{n/2}}\;.
\end{equation}
The zero crossing of the skewness, $B_3(\betapc) = 0$, determines the pseudo-critical 
$\betapc$ from the initial $\beta$-scans per fixed $(am,\ns,\nt)$.
Since there is only a small number of simulation points, reweighting with the multi-histogram method~\cite{Ferrenberg:1989ui} is 
used to interpolate and obtain a precise value for $\betapc$.
Repeating this for different masses provides the phase boundary as a function of mass,
$\betapc(am)$ for that volume. The kurtosis $B_4(\betapc)$ is evaluated along
this line, again employing the multi-histogram method to interpolate.

\begin{figure}
    \centering
    \includegraphics{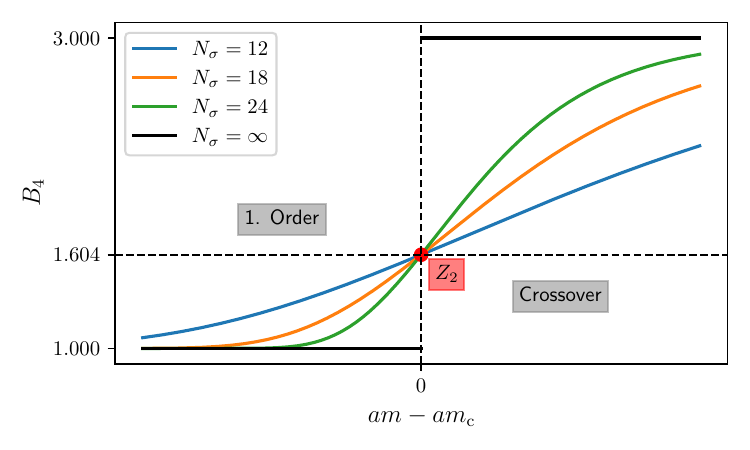}
    \caption{Schematic volume dependence of the 
    kurtosis $B_4$ near a $\ztwo$-critical point.}
    \label{fig:schematic-b4}
\end{figure}

Once this is available on three volumes, a finite-size scaling formula~\cite{Jin:2017jjp,Takeda:2016vfj}
\begin{equation}
	\label{equ:kurtosis-fss}
	B_4(\betapc;am,\ns)\approx\left(1.6044 + c(am-\amc)\ns^{1/\nu}\right)\left(1+b\ns^{y_t-y_h}\right)\;,
\end{equation}
is fitted to the kurtosis values for different masses $am$ and volumes $\ns$ in order to determine the critical quark mass $\amc$.
A schematic plot showing the volume dependence of $B_4$ in the vicinity of the $\ztwo$-critical 
point is shown in figure~\ref{fig:schematic-b4}.
The fit parameters $c$ and $b$ are arbitrary while $y_t=1/\nu=1.5870(10)$ and $y_h=\beta\delta/\nu=2.4818(3)$~\cite{Pelissetto:2000ek} are the universal 3D Ising exponents
governing the approach to the infinite spatial volume limit,
where the kurtosis turns into a step function from 1, representing a first-order transition, to 3 representing a crossover.
For sufficiently large volumes, the kurtosis lines corresponding to different volumes have a common crossing point at the 
universal 3D Ising value of $B_4(\amc)=1.6044$~\cite{Pelissetto:2000ek}.
When applying the kurtosis finite-size scaling formula for $\nf=5$ and $\nt\in\{4,6\}$
data, no correction term is needed, as it becomes insignificant when included in 
the fit.
For comparability, the kurtosis fits applied to distributions, which were generated by 
the ML model, also do not include the correction term.
The error of the critical mass is determined as the square root of the variance of the corresponding fit parameter as
given by the fit routine.

\section{Masked autoregressive flows}
The task of learning an unknown joint probability distribution $p(\vec{x})$ from data samples $\{\vec{x}\}$, where  $x_i,\; i=1,\ldots ,D$ denote a collection of $D$ observables, each observable representing a dimension in this space, is of general interest and can be formulated in several ways.
One natural way to formulate it is by using the chain-rule of probability and decomposing the joint density as a product over conditionals,
\begin{equation} \label{eq:autoreg}
    p(x_1,x_2, \hdots, x_D) = \prod^D_{i=1} p(x_i\mid \bm{x}_{1:i-1 }) \, ,
\end{equation}
with $p(x_1 \mid x_0) \equiv p(x_1)$ and where our choice of ordering reflects the fact that the probability of each dimension $x_i$ depends only on the previous elements in the $D$-dimensional vector, $\bm{x}_{1:j<i}$. Note here that $\bm{x}_{1:i-1}$ is shorthand for the vector of inputs corresponding to $(x_1, x_2 \cdots x_{i-1})$ and excluding $x_k$ with $k \geq i$. This sequential construction of the joint probability is commonly known as the \emph{autoregressive} property. While this autoregressive decomposition provides a general and exact representation of the joint distribution, its sequential nature introduces practical limitations when implemented in neural network architectures. Specifically, such types of models are known to have two shortcomings \cite{NIPS1999_e6384711}: \emph{(1)} for a $D$-dimensional space, a sequence of $D$ computations is required in order to reproduce the joint probability density, which renders parallelisation of such a model difficult; \emph{(2)} the models are sensitive to the order in which the conditionals appear. For a given network this means that some orderings of the input dimensions can reproduce the density more accurately than others.

In this work we employ a ML-based-autoregressive ansatz that tackles both of these shortcomings. The first of these shortcomings has been addressed in~\refer\cite{Germain:2015yft}, where the authors use the autoregressive property to propose a novel way to model each of the conditionals appearing in \eq\eqref{eq:autoreg}, using an autoencoder network\footnote{Autoencoder networks were originally introduced for constructing a lower dimensional representation \cite{Rumelhart:1986gxv,Hinton:2006tev} of high dimensional data using an encoder, a transformation that can then be reversed by a decoder to faithfully represent the original data. Such a task necessarily relies on the correlations present in the original data set, which in general holds for physical observables coming from lattice simulations.}, called the \emph{masked autoencoder} (MADE). This approach trades $D$ sequential computations for a single pass through all data, which can be performed in parallel, by introducing \emph{masks} in a fully connected network as shown schematically in figure~\ref{fig:MAF}. These masks selectively remove connections between input and output nodes, ensuring that each output $x_i$ depends only on the preceding inputs $x_{j<i}$, thereby enforcing the autoregressive structure.
Subsequently, a further improvement of this proposal was put forward in \refer\cite{Papamakarios:2017tec}, where the authors combined different numbers of MADE blocks to form an \emph{expressive flow}\footnote{A model is said to be expressive when it is able to capture all the distinguishing features of an underlying distribution, and the interpretation of this model as a (normalising) flow will be explained in the following section.} model called masked autoregressive flow (MAF). By joining these \emph{order-sensitive} MADE blocks in a chain, and permuting the order of inputs between each set of consecutive MADE blocks, as shown in figure~\ref{fig:MAF}, the second shortcoming listed above can be addressed, i.e., the MAF architecture implements an \emph{order-agnostic} autoregressive flow model. 

\subsection{Architecture and training protocol}\label{subsec:gen_arch}
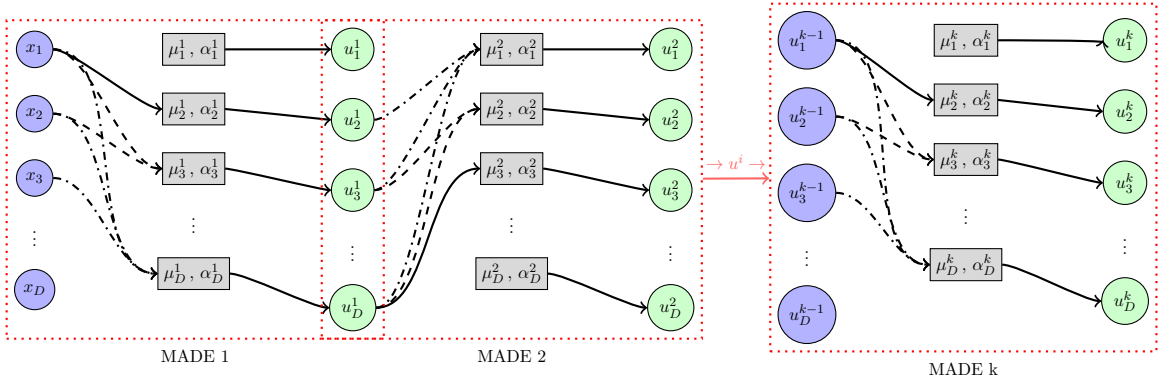
\begin{figure}
\centering
\begin{tikzpicture}[
    scale=0.60, transform shape, 
    node distance=1.4cm and 1.6cm,
    every node/.style={draw, minimum size=7mm, align=center},
    xnode/.style={circle, fill=blue!30},
    hidden/.style={rectangle, fill=gray!30},
    output/.style={circle, fill=green!20},
    arrow/.style={->, thick}
]

\node[xnode] (x1a) at (0.1,0.8) {$x_1$};
\node[xnode, below=0.6cm of x1a] (x2a) {$x_2$};
\node[xnode, below=0.6cm of x2a] (x3a) {$x_3$};
\node[xnode, below=1.6cm of x3a] (xDa) {$x_D$};

\node[draw=none] at ($(x3a)!0.5!(xDa)$) {$\vdots$};

\node[hidden] (h1a) at ($(x1a)+(3.5cm,0cm)$) {$\mu^1_1 \,,\, \alpha^1_1$};
\node[hidden, below=0.6cm of h1a] (h2a) {$\mu^1_2 \,,\, \alpha^1_2$};
\node[hidden, below=0.6cm of h2a] (h3a) {$\mu^1_3 \,,\, \alpha^1_3$};
\node[hidden, below=1.6cm of h3a] (hDa) {$\mu^1_D \,,\, \alpha^1_D$};

\node[draw=none] at ($(h3a)!0.5!(hDa)$) {$\vdots$};

\node[output] (o1a) at ($(h1a)+(3.5cm,0cm)$) {$u^1_1$};
\node[output, below=0.6cm of o1a] (o2a) {$u^1_2$};
\node[output, below=0.6cm of o2a] (o3a) {$u^1_3$};
\node[output, below=1.6cm of o3a] (oDa) {$u^1_D$};

\node[draw=none] at ($(o3a)!0.5!(oDa)$) {$\vdots$};

\draw[arrow] (x1a) to[out=0, in=180, looseness=0.4] (h2a);

\foreach \x in {x1a,x2a}
    \draw[arrow, dashed] (\x) to[out=0, in=180, looseness=0.6] (h3a);

\foreach \x in {x1a,x2a,x3a}
    \draw[arrow, dash dot] (\x) to[out=0, in=180, looseness=0.8] (hDa);

\draw[arrow] (h1a) to[out=0, in=180, looseness=0] (o1a);
\draw[arrow] (h2a) to[out=0, in=180, looseness=0] (o2a);
\draw[arrow] (h3a) to[out=0, in=180, looseness=0.2] (o3a);
\draw[arrow] (hDa) to[out=0, in=180, looseness=0.4] (oDa);

\begin{scope}[xshift=8cm]
\node[hidden] (h1b) at ($(o1a)+(3.5cm,0cm)$) {$\mu^2_1 \,,\, \alpha^2_1$};
\node[hidden, below=0.6cm of h1b] (h2b) {$\mu^2_2 \,,\, \alpha^2_2$};
\node[hidden, below=0.6cm of h2b] (h3b) {$\mu^2_3 \,,\, \alpha^2_3$};
\node[hidden, below=1.6cm of h3b] (hDb) {$\mu^2_D \,,\, \alpha^2_D$};

\node[draw=none] at ($(h3b)!0.5!(hDb)$) {$\vdots$};

\node[output] (o1b) at ($(h1b)+(3.5cm,0cm)$) {$u^2_1$};
\node[output, below=0.6cm of o1b] (o2b) {$u^2_2$};
\node[output, below=0.6cm of o2b] (o3b) {$u^2_3$};
\node[output, below=1.6cm of o3b] (oDb) {$u^2_D$};

\node[draw=none] at ($(o3b)!0.5!(oDb)$) {$\vdots$};

\draw[arrow] (oDa) to[out=0, in=180, looseness=0.8] (h3b);

\foreach \x in {o3a,oDa}
    \draw[arrow, dashed] (\x) to[out=0, in=180, looseness=0.6] (h2b);

\foreach \x in {o2a,o3a,oDa}
    \draw[arrow, dash dot] (\x) to[out=0, in=180, looseness=0.4] (h1b);

\draw[arrow] (h1b) to[out=0, in=180, looseness=0] (o1b);
\draw[arrow] (h2b) to[out=0, in=180, looseness=0] (o2b);
\draw[arrow] (h3b) to[out=0, in=180, looseness=0.2] (o3b);
\draw[arrow] (hDb) to[out=0, in=180, looseness=0.4] (oDb);
\end{scope}

\begin{scope}[xshift=17cm]
\node[xnode] (x1c) at (0.1,1.0) {$u^{k-1}_1$};
\node[xnode, below=0.4cm of x1c] (x2c) {$u^{k-1}_2$};
\node[xnode, below=0.4cm of x2c] (x3c) {$u^{k-1}_3$};
\node[xnode, below=1.4cm of x3c] (xDc) {$u^{k-1}_D$};

\node[draw=none] at ($(x3c)!0.5!(xDc)$) {$\vdots$};

\node[hidden] (h1c) at ($(x1c)+(3.5cm,0cm)$) {$\mu^k_1 \,,\, \alpha^k_1$};
\node[hidden, below=0.6cm of h1c] (h2c) {$\mu^k_2 \,,\, \alpha^k_2$};
\node[hidden, below=0.6cm of h2c] (h3c) {$\mu^k_3 \,,\, \alpha^k_3$};
\node[hidden, below=1.6cm of h3c] (hDc) {$\mu^k_D \,,\, \alpha^k_D$};

\node[draw=none] at ($(h3c)!0.5!(hDc)$) {$\vdots$};

\node[output] (o1c) at ($(h1c)+(3.5cm,0cm)$) {$u^k_1$};
\node[output, below=0.6cm of o1c] (o2c) {$u^k_2$};
\node[output, below=0.6cm of o2c] (o3c) {$u^k_3$};
\node[output, below=1.6cm of o3c] (oDc) {$u^k_D$};

\node[draw=none] at ($(o3c)!0.5!(oDc)$) {$\vdots$};

\draw[arrow] (x1c) to[out=0, in=180, looseness=0.4] (h2c);

\foreach \x in {x1c,x2c}
    \draw[arrow, dashed] (\x) to[out=0, in=180, looseness=0.6] (h3c);

\foreach \x in {x1c,x2c,x3c}
    \draw[arrow, dash dot] (\x) to[out=0, in=180, looseness=0.8] (hDc);

\draw[arrow] (h1c) to[out=0, in=180, looseness=0] (o1c);
\draw[arrow] (h2c) to[out=0, in=180, looseness=0] (o2c);
\draw[arrow] (h3c) to[out=0, in=180, looseness=0.2] (o3c);
\draw[arrow] (hDc) to[out=0, in=180, looseness=0.4] (oDc);
\end{scope}

\node[draw, red, dotted, thick, fit=(x1a) (xDa) (h1a) (o1a) (oDa),
      inner sep=1mm, label={[align=center]below:MADE 1}] (box1) {};
\node[draw, red, dotted, thick, fit=(o1a) (oDa) (h1b) (oDb),
      inner sep=1mm, label={[align=center]below:MADE 2}] (box2) {};
\node[draw, red, dotted, thick, fit=(x1c) (xDc) (h1c) (o1c) (oDc),
      inner sep=1mm, label={[align=center]below:MADE k}] (box3) {};

\draw[->, red!60, thick] (box2) -- node[above, draw=none]{$\to u^i \to$} (box3);
\end{tikzpicture}
\caption{Schematic diagram of the architecture of a typical masked autoregressive flow network with $D$-dimensional input vector $\bm{x}$ and $k$ MADE blocks. The blocks denote a chain of compositions of functions learned in each MADE block, $\mu^i_j(\bm{x}_{1:j-1})$ and $ \alpha^i_j(\bm{x}_{1:j-1})$. Between successive MADE blocks, permutations of the input vector are applied, modifying the autoregressive factorisation and enabling order-agnostic learning. Note that the grey boxes represent the hidden units of the NN: we have chosen to label them by the functions given in \eq(\ref{eq:hiddenFig2}) for better understanding as to what the NN learns.}
\label{fig:MAF}
\end{figure}

The architecture of the MAF model can be interpreted as a normalising flow, as we now explain. A normalising flow (NF) \cite{Rezende:2015ocs} is a neural-network-based framework for modelling complex target distributions by transforming a simple prior distribution (e.g. a standard normal) through a bijective and differentiable map. This allows both efficient sampling and exact evaluation of probability densities via the change-of-variables formula.
NFs can be trained in two settings: either from samples drawn from an unknown target distribution, or from a known distribution that is difficult to sample from directly. In both cases, the goal is to learn an invertible transformation that maps samples from a simple prior to the target distribution.
In this work, we consider the former setting, where samples from the target theory are obtained from lattice simulations. Our aim is to learn the underlying probability distribution and subsequently generate new samples using the trained flow. The appropriate training objective is maximum likelihood estimation, i.e. maximising the likelihood of the observed data under the model. For a transformation
$x=f(u)$, the likelihood is given by
\begin{align}\label{eq:lossNF}
    p_{X}(\bm{x}) = p_{\mathcal{N}(0,1)}\left(f^{-1}(\bm{x})\right) \left| \frac{\partial f^{-1}(\bm{x})}{\partial \bm{x}} \right|\;,
\end{align}
where this expression gives the probability density of a sample $x$ under the distribution $p_X$, by mapping $x$ to the normal distribution $\mathcal{N}(0,1)$ via the inverse transformation $f^{-1}$, scaled by the Jacobian of the transformation.
In other words, training is performed via the inverse mapping $u = f^{-1}(x)$, where we first try to learn the function $f^{-1} : x \to u$, which transforms samples from the target distribution into independent standard normal variables. Once learned, the forward map $f$ can be used to generate new samples from the target distribution. \\
As mentioned above, MAF consists of a sequence of MADE blocks. Each block employs an autoregressive parametrisation in which each dimension $x_i$ is modelled by a conditional Gaussian distribution whose parameters only depend on the preceding components of the $D$-dimensional input vector
\begin{align}\label{eq:lossMADE}
    p(x_i \mid \bm{x}_{1:i-1}) = \mathcal{N}(x_i \mid \mu_i, \exp(\alpha_i))\;.
\end{align}
Here, the transformation of the variables is defined as,
\begin{align}
    x_i = u_i \cdot e^{\alpha_i} + \mu_i\;,
\end{align}
with the shift and log-scale functions defined as
\begin{equation}\label{eq:hiddenFig2}
\mu_i = f_{\mu_i}(\bm{x}_{1:i-1}), \quad
\alpha_i = f_{\alpha_i}(\bm{x}_{1:i-1})\;.
\end{equation}
These parameters define an invertible affine autoregressive transformation, with the inverse recovered as
\begin{equation}\label{eqn:trans_rule}
u_i = \frac{x_i - \mu_i}{\exp(\alpha_i)} \;.
\end{equation}
Each MADE block therefore defines an invertible mapping
$f_k : \mathbb{R}^D \rightarrow \mathbb{R}^D$.
Stacking multiple MADE blocks results in a composite transformation
\begin{equation}
u = f_K \circ f_{K-1} \circ \dots \circ f_1 (x)\;,
\end{equation}
where, as mentioned before, the final latent variable $u$ is constrained to follow a standard multivariate normal distribution, $u \sim \mathcal{N}(0, I)$. A schematic illustration of this network can be found in figure~\ref{fig:MAF}.

The likelihood of the data is then evaluated using the change-of-variables
formula \eq(\ref{eq:lossNF}), with the Jacobian determinant of each affine autoregressive transformation contributing additively to the log-likelihood. The training procedure in the model then proceeds by driving the neural network (NN) parameters to those values that minimise the negative of this log-likelihood. To sample from the learned distribution, one must pass through the model in reverse order.

\subsection{Application to lattice data}\label{subsec:our_arch}

\begin{figure}
\centering
\begin{tikzpicture}[
    scale=0.73, transform shape, 
    node distance=1.2cm and 1.4cm,
    every node/.style={draw, minimum size=6mm, align=center},
    xnode/.style={circle, fill=blue!30},
    ynode/.style={rectangle, fill=orange!30},
    hidden/.style={rectangle, fill=gray!30},
    output/.style={circle, fill=green!20},
    arrow/.style={->, thick}
]

\node[xnode] (x1a) at (0.1,0.8) {$S$};
\node[xnode] (x2a) at (0.1,-1.2) {$\bar{\psi}\psi$};

\node[ynode] (y1) at (0.1,-3.5) {
    $y = \begin{pmatrix}
    am\\
    \beta\\
    \ns
    \end{pmatrix}$
};

\node[hidden] (h1a) at ($(x1a)+(2.5cm,1cm)$) {$h^1_1$};
\node[hidden, below=0.8cm of h1a] (h2a) {$h^1_2$};
\node[hidden, below=1.8cm of h2a] (hNa) {$h^1_N$};

\node[draw=none] at ($(h2a)!0.5!(hNa)$) {$\vdots$};

\node[output] (o1a) at ($(h1a)+(3cm,-1.5cm)$) {$p_1(S\mid y)$};
\node[output, below=of o1a] (o2a) {$p_1(\bar{\psi}\psi\mid S,y)$};

\node[draw=none] at ($(o1a)!0.45!(o2a)$) {$\times$};

\foreach \h in {h1a,h2a,hNa}
  \draw[arrow, dashed] (y1.east) to[out=0, in=180, looseness=0.4] (\h.west);
\foreach \h in {h1a,h2a,hNa}
  \draw[arrow] (x1a) -- (\h);
\foreach \h in {h1a,h2a,hNa}
  \draw[arrow] (\h) -- (o2a);

\begin{scope}[xshift=12cm]
\node[xnode] (x1b) at (0.1,0.8) {$u^1_1$};
\node[xnode] (x2b) at (0.1,-1.2) {$u^1_2$};

\node[ynode] (yb) at (0.1,-3.5) {
    $y=\begin{pmatrix}
    am\\
    \beta\\
    \ns
    \end{pmatrix}$
};

\node[hidden] (h1b) at ($(x1b)+(2.5cm,1cm)$) {$h^2_1$};
\node[hidden, below=0.8cm of h1b] (h2b) {$h^2_2$};
\node[hidden, below=1.8cm of h2b] (hNb) {$h^2_N$};

\node[draw=none] at ($(h2b)!0.5!(hNb)$) {$\vdots$};

\node[output] (o1b) at ($(h1b)+(3cm,-1.5cm)$) {$p_2(u^1_1\mid u^1_2,y)$};
\node[output, below=of o1b] (o2b) {$p_2(u^1_2\mid y)$};
\node[draw=none] at ($(o1b)!0.55!(o2b)$) {$\times$};

\foreach \h in {h1b,h2b,hNb}
  \draw[arrow, dashed] (yb.east) to[out=0, in=180, looseness=0.4] (\h.west);
\foreach \h in {h1b,h2b,hNb}
  \draw[arrow] (x2b) -- (\h);
\foreach \h in {h1b,h2b,hNb}
  \draw[arrow] (\h) -- (o1b);
\end{scope}

\node[draw, red, dotted, thick, fit=(x1a) (x2a) (y1) (h1a) (o2a),
      inner sep=2mm, label={[align=center]below:MADE 1}] (box1) {};
\node[draw, red, dotted, thick, fit=(x1b) (x2b) (yb) (h1b) (o1b),
      inner sep=2mm, label={[align=center]below:MADE 2}] (box2) {};

\draw[->, red!60, thick] (box1.east) -- node[above, draw=none]{
$u^1 = \begin{bmatrix}
    \frac{S - \mu_1(y) }{e^{\alpha_1(y)}}  \\ \\
    \frac{\bar{\psi}\psi - \mu_2(S,y)}{e^{\alpha_2(S,y)}}
\end{bmatrix}$
} (box2.west);
\end{tikzpicture}
\caption{Schematic of the MAF architecture, as applied to the lattice data, where the goal is to learn the conditional probability distribution $p(S, \bar{\psi}\psi\mid am, \beta, \ns)$ from true samples of $S$ and $\bar{\psi}\psi$ obtained from lattice simulations at fixed $am, \beta$ and $\ns$. The conditional probabilities associated with the first MADE block are implied by its autoregressive structure but are not propagated to subsequent blocks; they are shown only to illustrate the distribution that would be obtained if the flow were truncated after a single MADE block. The transformations given by \eq\eqref{eqn:trans_rule} are propagated between the blocks and are shown in red.}
\label{fig:ourMAF}
\end{figure}
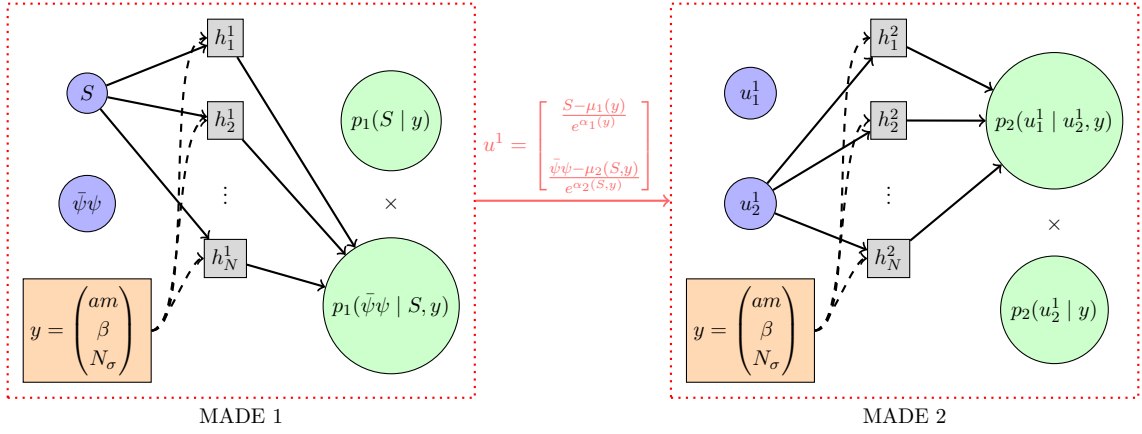

To connect this section to our data analysis, we now adapt the formalism described above to the lattice case study described in section~\ref{sec:latt_use_case}. We aim to learn the joint probability distribution $p(\pbp, S \mid \beta,  am, \ns)$ of the action $S$ and chiral condensate $\bar{\psi}\psi$, i.e.~$D=2$, conditioned on the spatial lattice extent $\ns$, the fermion mass $a m$ and the lattice gauge coupling $\beta$. An important difference from the discussion in section~\ref{subsec:gen_arch} is the introduction of \emph{external} conditional variables $\bm{y}$. The two sets of inputs enter differently: the connections carrying $\bm{x}_{1:i-1}$ are masked so as to enforce the autoregressive property, whereas the connections carrying $\bm{y}$ are left unmasked, so that every hidden unit of every MADE block has access to the full vector $\bm{y}$. In order to explain the NN modified for this case, we refer the reader to figure~\ref{fig:ourMAF}, where the conditional input denoted by $\bm{y}$ is fed into the hidden layer of every MADE block in the chain. These conditional parameters enter the training procedure via the learnable functions $\mu_i$ and $\alpha_i$, which are no longer just dependent on the previous input dimensions but also on the parameters $\bm{y}$ via
\begin{align}\label{eq:cond_dep}
\mu_i &= f_{\mu_i}(\bm{x}_{1:i-1} ,\, \bm{y})\;, \nonumber \\
\alpha_i &= f_{\alpha_i}(\bm{x}_{1:i-1} ,\, \bm{y})\;,
\end{align}
where the functions belonging to the first dimension (with label $i=1$) depend only on $\bm{y}$. The functions $f_{\mu_i}$ and $f_{\alpha_i}$ of Eq.~\eqref{eq:cond_dep} are therefore ordinary feed-forward networks of the hidden layers and units in each layer, along with their activations, taking the concatenated vector ($\bm{x}_{1:i-1}, \bm{y}$) as input, with the components of $\bm{y}$ standardized to zero mean and unit variance over the training set. No separate interpolation scheme is applied: the dependence of the density on $\bm{y}$ is implicit in the learned function itself and can be evaluated at any desired $\bm{y}$.
In figure~\ref{fig:ourMAF}, we depict our model for the case with a single MADE block with $N$ hidden units, where we can directly interpret the outputs as the desired probability distribution $p(\pbp, S \mid \beta,  a m, \ns) = p_1(S \mid \beta,  a m, \ns) \times p_1(\bar{\psi}\psi \mid S,\beta,  a m, \ns)$, and for the case of more than one MADE block, we show the quantities passed along the MAF chain. The number of hidden units $N$ and the number of MADE blocks determine the complexity of the MAF model, which is ultimately determined by the complexity of the target distribution we aim to learn\footnote{Appendix~\ref{sec:AppB} contains a discussion on how the quality of learned complex distributions depends on the number of MADE blocks and hidden units $N$, using the Gaussian mixture model (GMM) as our test case.}.

\subsection{Sources for uncertainties: statistics vs systematics}
ML-based analyses are affected by different sources of uncertainty. In the following, we distinguish between two types of uncertainty, which we refer to as \emph{statistical} and \emph{systematic}. 
By \emph{statistical uncertainty} we refer to the error associated with estimating observables from a finite number of samples drawn from a \emph{fixed trained model}. This refers to the case where we fix the training seed, thereby fixing the initialisation of weights, and sources of explicit randomness like data shuffling\footnote{Note that fixing the training seed does not guarantee identical output weights since training can still differ based on the underlying hardware.}. This procedure greatly reduces variation in the trained model. Hence the uncertainty coming from this type of training is essentially the standard error of the estimator and can be reduced arbitrarily by increasing the number of generated samples. Therefore, this sampling error does not represent a faithful estimate of the uncertainties originating from the distribution of the training data.
Hence, we consider another contribution which we call the \emph{systematic uncertainty} arising from the dependence of the results on different trained ML models, i.e., obtained by changing the training seed. Note that, throughout this work, the term \enquote{systematic uncertainty} is applied exclusively to contributions introduced during the training process for a fixed ML model. We do not systematically investigate uncertainties arising from the choice of ML architecture, hyper-parameters, or other aspects of model design. Instead we have optimized the hyper-parameters to best reproduce the lattice data as discussed in appendix~\ref{sec:appendixNNparams} and \ref{sec:AppB}.

We emphasise that this systematic contribution cannot be reduced by increasing the number of samples from a single model, since there is a bias intrinsic to each trained model. To address this bias, one must instead consider multiple trained models\footnote{We further emphasise that the terminology and distinction used in this work, are specific to ML-based analyses.}. Since the models are trained on lattice data, the source of the systematic uncertainty ultimately originates from the statistical fluctuations and correlations present in the underlying data set. Moreover, the observed run-to-run variability reflects the interplay between the finite amount of lattice data available for training and the non-uniqueness of the learned representation. 
While the loss function, as in our case, indicates convergence to very similar values across independent training runs (see, for example figure~\ref{fig:LossNt6runs} and the discussion in appendix~\ref{sec:appendixNNparams}), these differences lead to slightly different learned distributions, which in turn propagate to the observables and produce a finite spread across runs.
 
In this work, the intra-model sampling error is kept negligible by generating sufficiently many samples per model. We take the variation across independent training runs as our primary estimate of the uncertainty of the ML-based observables, noting that in an idealised limit of infinite training data and a unique global optimum for the training objective, this run-to-run variability would vanish. For concreteness, we describe the two sources of uncertainty present in our analysis with explicit examples below.
 
\subsubsection{Sampling error from one model}
\label{sec:sampling_error_one_model}
In what follows, for a fixed trained ML model, we estimate observables by drawing $N_{\mathrm{samp}} = 10^6$ independent and identically distributed (i.i.d.) samples from the learned probability distribution. Assuming that the standard deviation of an observable $O$ is denoted by $\sigma_O$, the statistical uncertainty of the estimator is then given by 
\begin{equation}
\delta_O = \frac{\sigma_O}{\sqrt{N_{\mathrm{samp}}}}\;.
\end{equation}
Since sampling from the trained model is computationally inexpensive\footnote{See appendix~\ref{sec:appendixNNparams}: approximately 0.5 seconds per $10^6$ samples at fixed $\beta$, mass, and $\ns$.}, this quantity can be made arbitrarily small by increasing $N_{\mathrm{samp}}$. 

In order to quantify this, we provide the estimates for the mean of the chiral condensate $\langle \pbp \rangle$, the corresponding $\sigma_O$ and $\delta_O$ for $1$M samples generated from a model trained on only one part of the data for $\nt=4$, $\ns=\{8,16\}$, thus leaving out $\ns=12$ evaluated at the critical values of $\beta = \betapc$ and $am=\amc$. These values of $\beta$ and $am$ are chosen to reflect the goal of our analysis, namely estimating the critical boundary from the ML-analysis. As can be seen from table~\ref{tab:StatErr}, sampling $1$M already corresponds to a relative error of roughly $\sim0.02\%$. This can be compared to the systematic variation (explained in the next section) coming from different runs in figure~\ref{fig:runs_sys_unv1}.

\begin{table}[h!]
\centering
\begin{tabular}{|c|c|c|c|}
\hline
  & $\langle \pbp \rangle$ & $\sigma_{\pbp}$ & $\delta_{\pbp }$ \\ \hline
$\ns=8$ & $0.995028$ & $0.175201$ & $0.000175$ \\ \hline
$\ns=12$ & $0.980604$ & $0.152166$ & $0.000152$ \\ \hline
$\ns=16$ & $0.995921$ & $0.124025$ & $0.000124$ \\ \hline
\end{tabular}
\caption{Estimates for typical uncertainties when trained with single seed: This table shows values for $\nt=4$ data, generated with a model trained on $\ns=\{8,16\}$, thus leaving out $ \ns=12$, evaluated at $\betapc=4.98304866$ and $\amc = 0.08206206$, using $1$M samples. }\label{tab:StatErr}
\end{table}

Before comparing statistical and systematic uncertainties, we emphasise that sampling from the ML model yields i.i.d.\ samples of the chiral condensate and the action, as we learn the joint distribution $p(\bar{\psi}\psi, S)$ at fixed $\beta, am$, and $ \ns$. However, higher-order cumulants are not sampled directly, but are instead computed as nonlinear functions of the chiral condensate. Hence, a reliable propagation of errors to higher-order cumulants is non-trivial, analogous to the situation encountered in lattice analyses.

\subsubsection{Variation of the model}
\label{sec:variation_of_model_ensemble}
The second source of uncertainty described above arises from the dependence of the results on the specific trained ML model. To quantify this effect, we adopt an ensemble approach by performing $N_{\mathrm{runs}}$ independent training runs of the ML model on the same lattice data set, using different random initialisations (training seeds).

For each trained model $r = 1, \dots, N_{\mathrm{runs}}$, we draw $N_{\mathrm{samp}} = 10^6$ i.i.d.\ samples and compute an estimator $\hat{O}_r$ for the observable of interest. The final estimate is obtained as the ensemble average
\begin{equation}\label{eq:mean_runs}
\bar{O} = \frac{1}{N_{\mathrm{runs}}} \sum_{r=1}^{N_{\mathrm{runs}}} \hat{O}_r,
\end{equation}
with an associated uncertainty defined as the standard deviation across independent runs,
\begin{equation} \label{eq:std_runs}
\delta O_{\text{runs}} 
= \left( \frac{1}{N_{\mathrm{runs}} - 1} 
\sum_{r=1}^{N_{\mathrm{runs}}} (\hat{O}_r - \bar{O})^2 
\right)^{1/2}.
\end{equation}

\begin{figure}[ht]
    \centering
    \includegraphics[width=1 \textwidth]{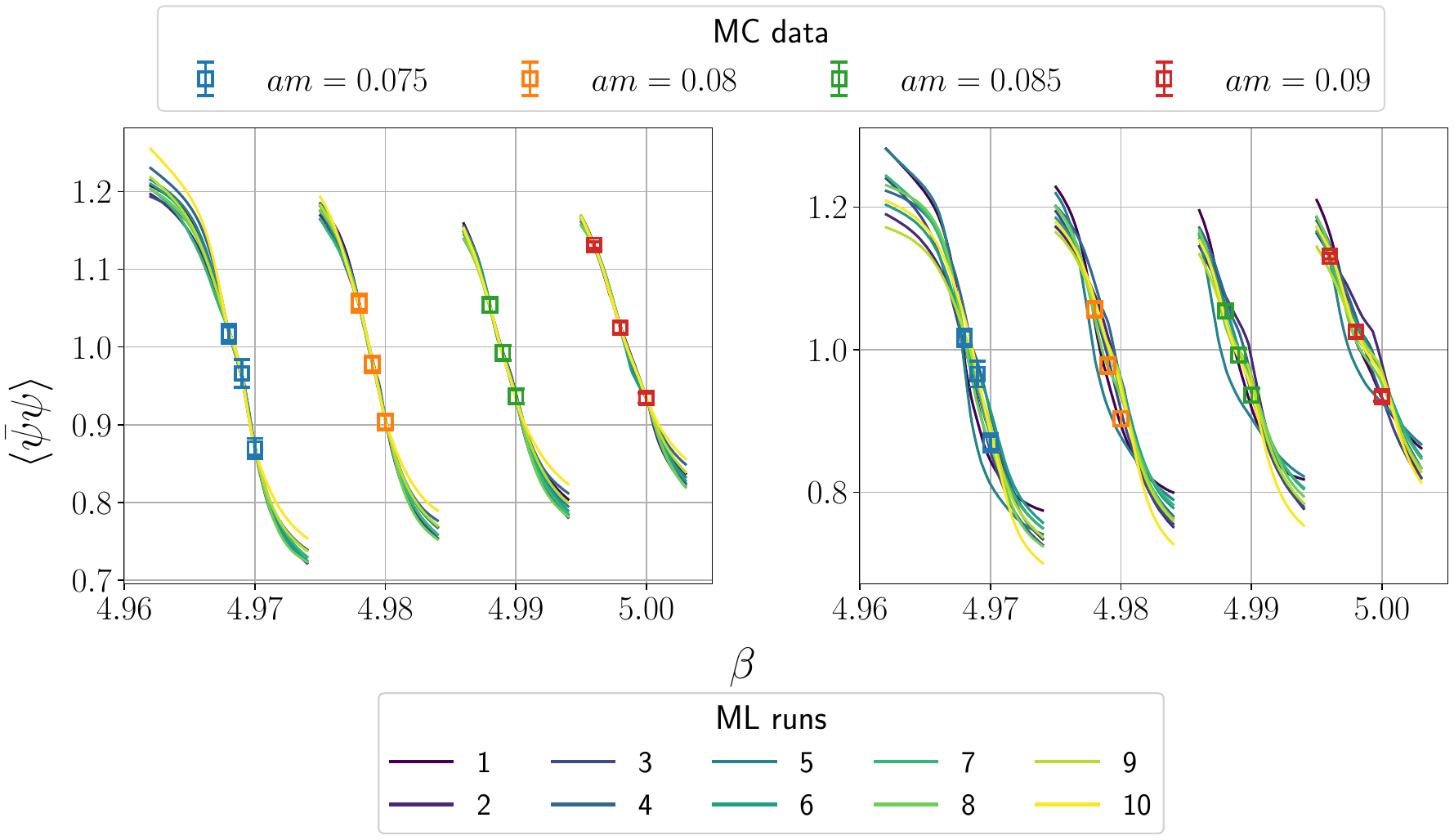}
\caption{ML predictions across independent training runs compared with MC data for $\nt=4$, $\ns=12$ data. Left: training on all data yields consistent results across runs. Right: excluding the $\ns=12$ data leads to increased run-to-run variation, reflecting reduced constraints on the learned distribution while maintaining consistency with the MC data.}
    \label{fig:runs_sys_unv1}
\end{figure}

This quantity measures the run-to-run variability of the ML model predictions and reflects the sensitivity of the learned distribution to the stochastic elements of the training procedure. In figure~\ref{fig:runs_sys_unv1}, we show the variation across independent training runs. The left panel corresponds to training on the full data set, where different runs agree closely. The right panel shows the case in which the $\ns=12$ data are excluded from training, leading to a visibly larger spread across runs. While the ML predictions remain consistent with the MC data in both cases, the increased variability in the right panel reflects the reduced constraint on the model when data are removed from the training set.

In our calculations with $N_{\mathrm{samp}}=10^6$, the sampling uncertainty is numerically negligible compared with $\delta O_{\mathrm{runs}}$. Consequently, the quoted uncertainty is dominated by run-to-run (model-to-model) variation and is effectively unchanged by further increasing $N_{\mathrm{samp}}$. Additionally, this systematic uncertainty can be applied directly to the higher-order cumulants, unlike the statistical uncertainty, as it is computed over the mean values of the cumulants obtained from independent runs. 

\section{ML-learned distributions vs MC distributions}\label{sec:learning-distr}
The MAF models were trained on MC data generated on lattices with temporal extents
$ \nt = \{4,6\} $ over a wide range of parameters $ \{\beta, am, \ns\} $, see table \ref{tab:Nf5_statistics}.
In the following, we assess how well the MAF models are able to
reproduce these data and discuss the extent to which the MAF can be used for interpolation within this parameter space.
We then investigate how accurately the critical mass marking the
$\ztwo$-boundary can be determined from the learned distributions using MAF models,
and assess the potential of this approach to reduce the number of required MC simulations.

\subsection{Testing the ML model: parameter interpolation}
In this section, we present results based on the MAF-learned density $p\left(\pbp, S\mid \beta, am, \ns \right)$ for the case $\nt=4$\footnote{Results of the cumulant analysis for $\nt = 6 $ are not shown in the main part of this manuscript, but in appendix~\ref{sec:complementary_nt6}. Qualitatively, they are very similar to those for $\nt=4$ and we include them in our discussion on the results of the critical mass estimation.}.
The quality of the learned distribution can be assessed either by comparing its cumulants, or by the direct comparison of the full distributions. To quantify the latter, we employ the \emph{maximum mean discrepancy} (MMD), which defines a ``distance'' between the learned distribution and that obtained from lattice simulations. We use this metric to compare the \emph{relative} learning of distributions, for the different parameter sets the data is conditioned on, and the results are shown in appendix~\ref{sec:appFigs}\footnote{We also use the MMD as a metric to quantify the training procedure for the GMM as discussed in detail in appendix~\ref{sec:AppB}.}. 
At the level of cumulants (\eq\eqref{equ:std-moments}), we restrict the comparison here to the mean value of the order parameter $\pbp$, the skewness $B_3\left( \pbp \right)$ and the kurtosis $B_4\left( \pbp \right)$ -- the quantities relevant in our analysis for determining the pseudo-critical $\beta_{\mathrm{pc}}$ and the critical quark mass $\amc$. For a full comparison, we refer the reader to appendix~\ref{sec:appFigs}, where results comparing the ML analysis with all available MC data are presented, including the second cumulant, namely the susceptibility $\chi(\pbp)$.

In the following, we test the model's ability to interpolate distributions
\emph{(1)} in the coupling constant $\beta$, \emph{(2)} in the mass $am$ and \emph{(3)} in the volume $\ns$.
Note that cases \emph{2} and \emph{3} are effectively interpolations in two dimensions as they require interpolation in the coupling constant as well.

\subsubsection{Interpolation in coupling}\label{sec:binter}

In this case the ML models were trained on the complete set of available MC simulation data obtained on lattices of size $\ns^3\times \nt$ either with temporal lattice extent $\nt = 4$ or $\nt=6$ (we will only show results for $\nt=4$).
In particular, data was provided for three spatial volumes, $\ns = \{8,12,16\}$. For each volume, four masses are available, and for each mass, three different coupling values $\beta$ were chosen to cover both sides of the transition point.
The critical mass for the $\ztwo$-boundary, determined as discussed in section~\ref{sec:estimating-ztwo} from this data set, was found to be $\amc \approx 0.082$~\cite{Cuteri:2021ikv}.
This chiral transition is then of first order for the smaller simulated mass values, $am = \{0.075,0.08 \}$, and an analytic crossover for the larger simulated masses, $am = \{0.085,0.09 \}$.

\begin{figure}[ht!]
    \centering
   \includegraphics[width=\textwidth]{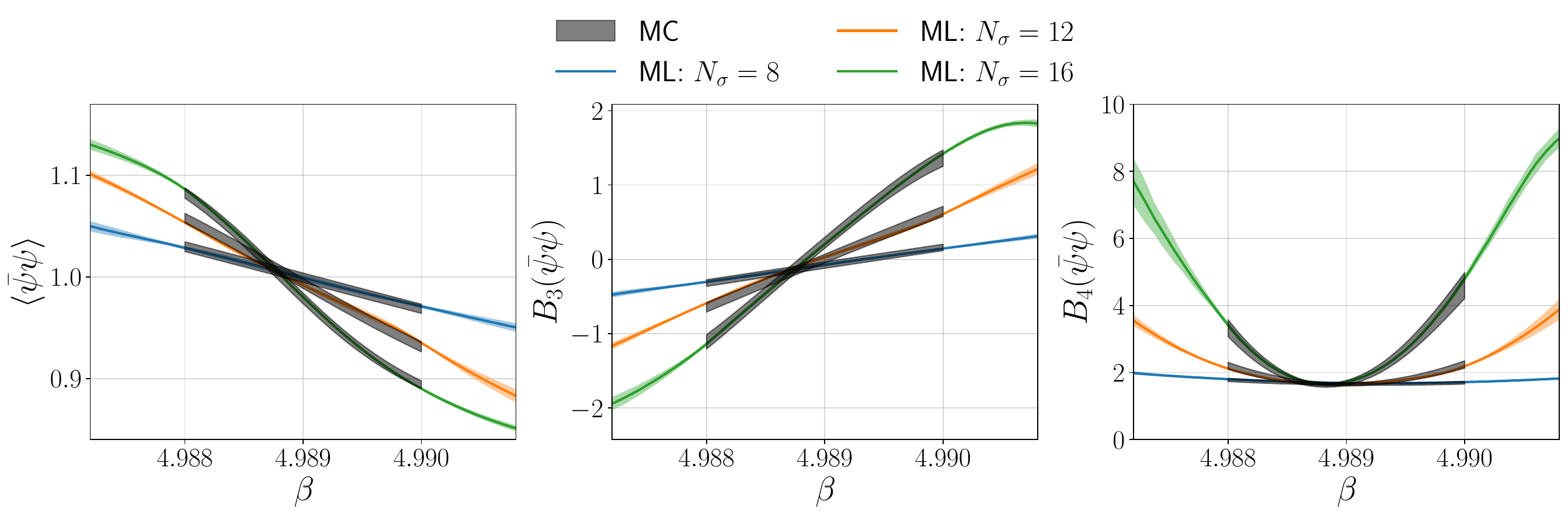}
    \caption{Interpolation in $\beta$ at $am=0.085$ and $\nt =4$. The mean $\langle  \pbp\rangle$, skewness $B_3\left(\pbp \right)$ and kurtosis $B_4\left(\pbp \right)$ (\eq\eqref{equ:std-moments}) of standard reweighted MC data (black) is compared against ML generated data for three different volumes.}
    \label{fig:beta_interpolation}
\end{figure}

Once the probability density $p\left(\pbp, S \mid \beta, am, \ns \right)$ has been learned, the model can be used to generate samples for arbitrary values of $\beta$.
Hence, this corresponds to a 1D interpolation in the coupling, comparable to the reweighting techniques~\cite{Ferrenberg:1989ui} commonly used in 
lattice analyses. In figure~\ref{fig:beta_interpolation}, we show a comparison of results for $\langle  \pbp\rangle$, $B_3\left(\pbp \right)$ and $B_4\left(\pbp \right)$ obtained from the MAF-learned distribution, from left to right, respectively, against standard $\beta$-reweighting. The coloured bands show the cumulants obtained from the ML model. The blue, orange and green bands, respectively, correspond to the $\beta$-interpolated results for $\ns = \{8,12,16\}$  for $am=0.085$, and should be compared with the MC values in black, obtained via standard reweighting. As can be seen from the figure, good agreement is obtained between the $\beta$-reweighted results and the ML-based results. The ML-based results lie within errors of the MC values of $\langle \pbp \rangle$ for all volumes. For better visualisation we only show results for $am=0.085$ here. The comparison of ML-based cumulants against MC data for other masses can be found in figure~\ref{fig:cumulantNt4_all} in appendix~\ref{sec:appFigs}. The error bands for the ML results show the standard deviation across runs, computed using \eq\eqref{eq:std_runs}, as discussed in section~\ref{sec:variation_of_model_ensemble}.
Based on figure~\ref{fig:beta_interpolation}, a further observation is that for parameter values outside the region covered by the MC training data, different ML runs begin to show increasing deviations, and the results exhibit growing fluctuations. This behaviour is expected for an extrapolation beyond the domain on which the model was trained. 
A more detailed evaluation of the variation of the probability distributions and the cumulants is given in appendix~\ref{sec:appFigs} via an MMD analysis.

\FloatBarrier

\subsubsection{Interpolation in mass}\label{sec:minter}

\begin{figure}[ht]
    \centering
    \includegraphics[width=\textwidth]{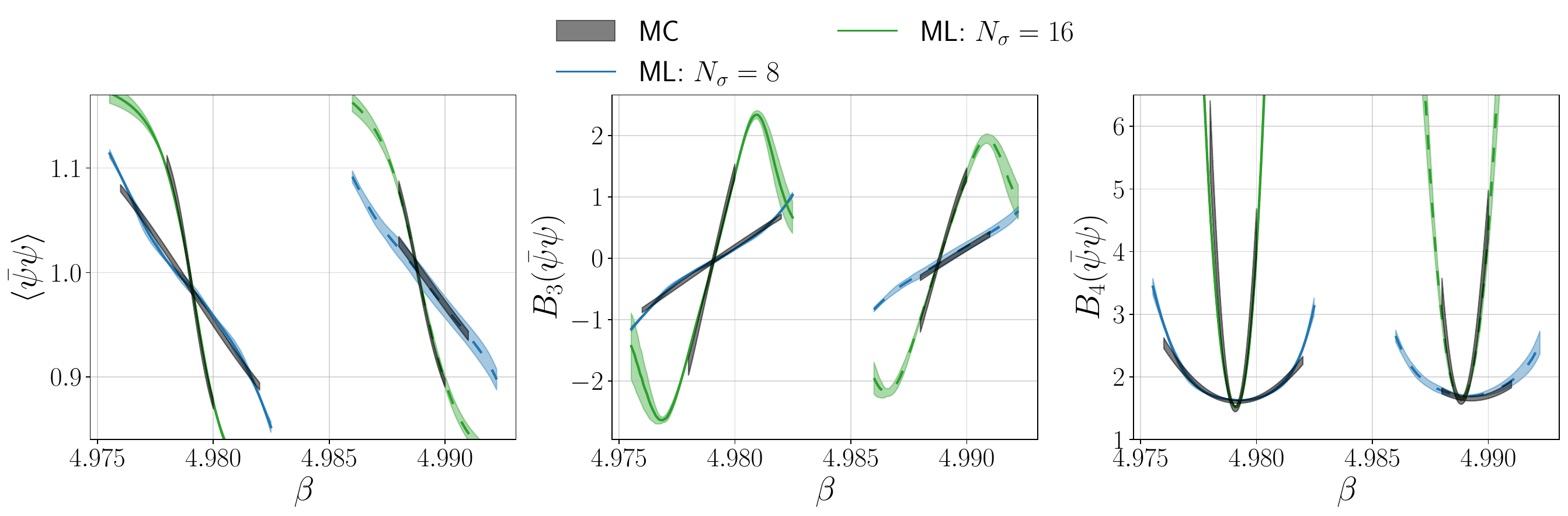}
      \caption{Interpolation in mass $am$ for $\nt=4$. The mean $\langle  \pbp\rangle$, skewness $B_3\left(\pbp \right)$, and kurtosis $B_4\left(\pbp \right)$ (\eq\eqref{equ:std-moments}) of standard reweighted MC data (black) are compared with ML-generated data for two different volumes. Solid lines show different volumes corresponding to mass $am=0.08$, which is (among others) included in the training data, dashed lines show $am=0.085$, which is obtained entirely through interpolation.} 
      \label{fig:mass_interpolation}
\end{figure}

While multi-histogram techniques can be adapted to also interpolate in mass,
this requires an accurate determination of the computationally expensive 
fermion determinant, and for this reason is rarely applied in practice.
A successful ML-based interpolation would thus constitute an economical alternative. \\
To test this capability, we used the same MC data set for $\nt = 4$, but deliberately excluded all data corresponding to one mass value, $am = 0.085$, across all $\ns$ during training. In figure~\ref{fig:mass_interpolation}, we compare the resulting $\langle \pbp \rangle$, $B_3(\pbp)$, and $B_4(\pbp)$ cumulants with both the MC data and the $\beta$-reweighted results from figure~\ref{fig:beta_interpolation}, for spatial lattice extents $\ns = {8,16}$. Results are shown for $am=0.08$ (solid lines), which is included in the training data, and for $am=0.085$, which is obtained entirely through interpolation.
A notable feature of the ML-based predictions at the interpolated mass $am=0.085$ is the visibly larger error bands, obtained from the standard deviation across multiple training runs (see \eq\eqref{eq:std_runs}), compared to those in figure~\ref{fig:beta_interpolation}. As discussed in section~\ref{sec:variation_of_model_ensemble}, this behavior is expected since the model has not been exposed to data at this mass value, leading to increased disagreement among ensemble members. Nevertheless, the predictions remain consistent with the lattice data at this mass, indicating that the model is capable of meaningful interpolation in mass and, consequently, of reducing the need for additional simulations.  
The extent to which this approach reproduces reliable estimates of the critical coupling will be examined in section~\ref{sec:z2-results}. We further refer the reader to figure~\ref{fig:cumulantNt4_remM085} for comparison of ML-based results with MC data for the remaining mass values and $\ns=12$ results and for the comparison at the level of distributions using MMD.

\FloatBarrier

\subsubsection{Interpolation in volume}\label{sec:vinter}

\begin{figure}[ht]
    \centering
    \includegraphics[width= \textwidth]{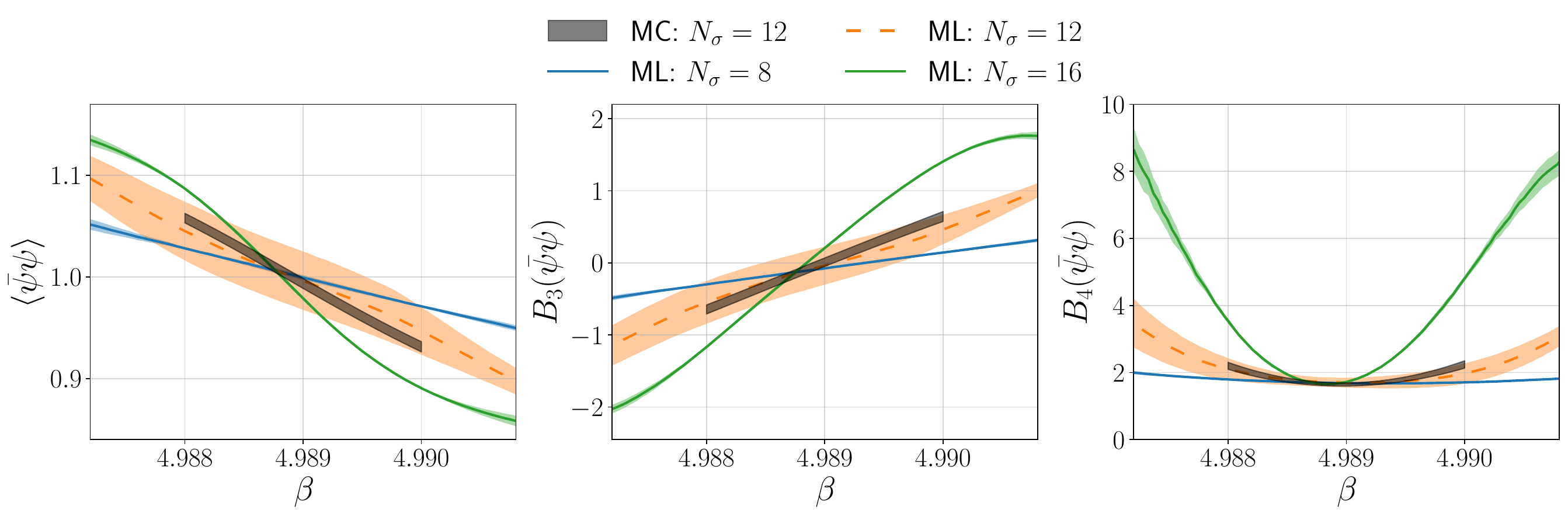}
    \caption{Interpolation in volume for $am=0.085$ and $\nt=4$. The mean $\langle  \pbp\rangle$, skewness $B_3\left(\pbp \right)$ and kurtosis $B_4\left(\pbp \right)$ (\eq\eqref{equ:std-moments}) of standard reweighted MC data (black) are compared with ML-generated data for three different volumes. Solid lines show the volumes $\ns=\{8,16\}$ which were included in the training data, the dashed line shows volume $\ns =12$ which is obtained entirely through interpolation.}
\label{fig:volume_interpolation_cumulants}
\end{figure}

Next we test the ability of our ML model to interpolate in volume,
for which no reweighting procedure based on MC histograms exists. 
For this purpose, all data corresponding to the intermediate spatial lattice extent $\ns=12$ were omitted from the  $\nt = 4$ MC data set during training. To assess the quality of the interpolated density, figure~\ref{fig:volume_interpolation_cumulants} shows the mean, the skewness and the kurtosis for a single mass, $am = 0.085$. The coloured bands show the cumulants obtained from the ML model. The orange band corresponds to the interpolated $\ns = 12$, and should be compared with the MC values in black, obtained via standard reweighting. It is apparent that the interpolation works not only at the level of the mean of our observable, but also for higher-order cumulants. For applications such as those discussed in 
section~\ref{sec:estimating-ztwo}, the zero crossing of the skewness, $B_3(\beta) = 0$ and the value of the kurtosis at this point are
of particular interest, as they characterise the location and the nature of the transition. 
Both are correctly reproduced within error bars at the interpolated volume $\ns=12$.
Good agreement between the $\langle \pbp \rangle$ values from the ML-learned
samples and the MC values can be seen.
Apparently, constraining the model with simulations at one smaller and one larger value of $\ns$ is sufficient for interpolation in $\ns$.
These results indicate that it may be possible to omit the full set of MC simulations for an intermediate volume, provided it lies within the range of simulated volumes. The computational resources saved in this way could instead be redirected towards increasing statistics at the largest volume.
At the same time, we observe that the standard deviation of the cumulants at the interpolated volume is noticeably larger than that obtained from direct MC simulations. For the results shown, this deviation is estimated from $10$ independent training runs. While increasing the number of runs can reduce the resulting uncertainties, the larger spread is expected, as the model has not been trained on data corresponding to $\ns = 12$. This leads to an increased uncertainty in derived quantities, such as the pseudo-critical $\beta$ or the kurtosis at the transition point.
Overall, these findings suggest that ML-based interpolation can serve as a useful tool for reducing computational cost, particularly in exploratory studies or when moderate precision is sufficient. However, for high-precision determinations, direct MC simulations remain essential.

\FloatBarrier
\subsection[Estimating the \texorpdfstring{$\ztwo$}{Z2}-boundary]{Estimating the \texorpdfstring{\boldmath$\ztwo$}{Z2}-boundary}\label{sec:z2-results}
\begin{figure}[h]
    \centering
    \includegraphics{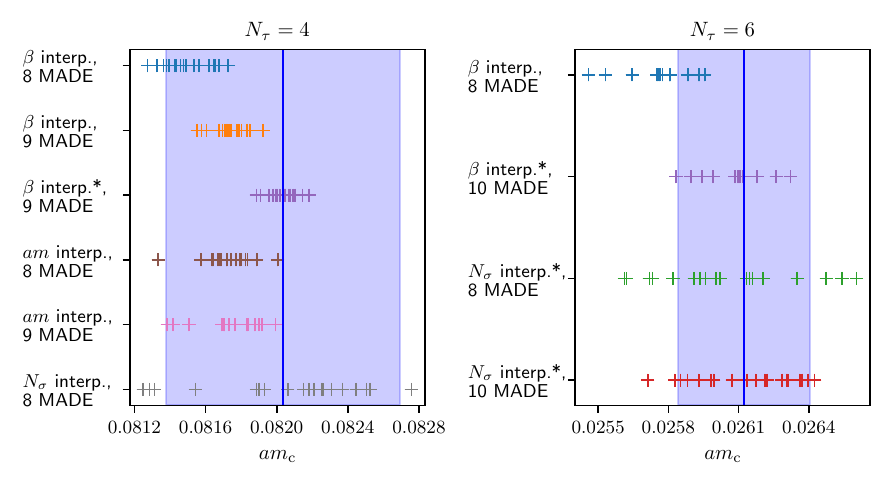}
    \caption{Comparison of results for the critical mass for $\nt=4$ (left) and $\nt=6$ (right) for the three test cases: $\beta$ interpolation, $am$ interpolation, and $\ns$ interpolation.
    The critical mass obtained directly from the lattice MC data is shown as a blue line with the error band in light blue. The asterisk (*) marks test cases, in which additional data was generated from the trained MAF models at mass values that were not simulated.}
    \label{fig:amc-comparison}
\end{figure}

In order to quantitatively assess the applicability of the ML model to
different training data sets, we evaluate the critical mass $\amc$ marking the $\ztwo$-boundary for each set and compare it with the critical mass obtained from the finite-size scaling
analysis of reweighted lattice data.
The procedure is the same for every test case:
We train the ML model on the given training set for a number of different random initialisations of the ML parameters, which are represented by different seeds.
For each seed we evaluate the skewness and kurtosis of the learned $\pbp$-distribution 
for a fixed set of parameter values $\{am, \ns, \beta\}$ within the range of the 
simulated values.
To determine the critical mass $\amc$ using the kurtosis finite-size scaling formula from \eq\eqref{equ:kurtosis-fss},
we follow the same steps as described in section~\ref{sec:estimating-ztwo}.
Note that the kurtosis values extracted from the learned distributions carry no meaningful uncertainty. As a result, the fit does not provide an uncertainty estimate for the critical mass $\amc$.
However, each seed yields a different critical mass resulting in a cloud of points, which gives a
rough estimate of the variability of the model with respect to different initial
parameters.
Finally we compare the cloud of critical masses obtained from ML to the critical mass value obtained from reweighted lattice data.
A graphical visualisation of the results is shown in figure~\ref{fig:amc-comparison} for both 
$\nt=4$ and $\nt=6$.

The first test case for both $\nt=4$ and $6$ uses all available data for training and employs a MAF model consisting of 8 MADE 
blocks. The corresponding critical masses are illustrated in blue in figure~\ref{fig:amc-comparison}, with $\nt=4$ shown on the left and $\nt=6$ on the right. The critical mass values obtained from ML show a systematic shift towards smaller values, which can be traced back to a so-called \emph{mode-covering} or \emph{bridging} effect when learning bimodal distributions \cite{Minka2005Divergence,Hackett:2021idh,Nicoli:2023qsl}:
Learned distributions with a two-peak structure representing a first-order transition display an unphysical weight in the gap between the two phases. This leads to larger kurtosis values, implying a shift of the critical quark mass to smaller values in the fit 
function, \eq\eqref{equ:kurtosis-fss}. We demonstrate this in appendix~\ref{sec:GMM_dist} through experiments using GMMs with widely separated means. Increasing the complexity of the MAF models helps to better capture the two-peak structures. In particular, increasing the number of MADE blocks from eight to nine reduces the deviation of the predicted critical masses from the MC results, as shown in orange.

The next test case, shown in purple, incorporates additional data generated by the ML model for mass values beyond those included in the original training set. The number of mass points was increased from four to eight for $\nt=4$ and from three to eight for $\nt=6$. For both $\nt=4$ and $6$, we observe a systematic shift towards larger critical mass values when the number of mass values is increased.  In both cases, the point cloud is scattered around the central value of the MC result. The systematic shift suggests the presence of systematic deviations in the learned distributions, the magnitude of which appears to depend on the quark mass.
By investigating the critical behaviour of the extracted kurtosis values
using the kurtosis finite-size scaling formula,
these systematic deviations are exposed whenever their magnitude varies with the quark mass
and the additionally generated data are unevenly distributed 
across the simulated mass range.

The results shown in brown for $\nt=4$ correspond to the case where one mass value is removed from the training data and interpolated using the ML model. The results exhibit only a slight increase in the spread of the critical mass values across different seeds, while the deviation from the MC results remains comparable to that in the case in which all mass values are included (blue). This suggests that interpolation in mass is reliable when a single mass value is omitted from a set of four. In this case, increasing the model complexity from eight to nine MADE blocks does not lead to a significant change in the results, as indicated by the pink points.

Lastly, the impact of removing the central volume from the set of simulated volumes is investigated. The ML predictions of the critical mass for $\nt=4$, shown in grey, exhibit a significantly larger variability than in the mass interpolation case, although no systematic deviation from the MC reference value is observed. As can be observed for $\nt=6$, increasing the model complexity to ten MADE blocks leads to a noticeable reduction in the spread (red). This is expected as MAF models with larger numbers of MADE blocks are able to learn more complex distributions.

The test cases shown in figure~\ref{fig:amc-comparison} demonstrate that interpolation in mass and in volume works quite well.
However, test cases involving extrapolation showed large shifts of the critical mass values, which we did not include in
figure~\ref{fig:amc-comparison} to maintain the readability of the other results.

As an example, removing $\ns=16$ and $\ns=18$ from the training data of $\nt=6$ leaves only $\ns=12$ for the smallest mass $am=0.02$, since MC simulations at this mass are available only for $\ns=12,18$. Any ML-based estimate of cumulants at $am=0.02$ therefore constitutes an extrapolation, regardless of the target volume.
This leads to critical mass values of $\amc=0.0230(9)$ which misses the reference value of $\amc=0.02612(28)$ by approximately $3.3\sigma$.

\subsection{Application to critical scaling}
Another advantage of the ML-approach is that the trained model can sample at the predicted values of $\amc$ and $\betac$ to explore 
the universal properties of the joint distribution, which go beyond the third- and fourth-order cumulants. In particular, the full shape of the distribution at a critical point, transformed to statistically independent coordinates (e.g. through a principal component analysis \cite{Hotelling1933PCA}), is universal \cite{Wilding_1992}. A simple parametrisation of this transformation is given in terms of energy-like $\mathcal{E} = S + r \cdot \bar{\psi}\psi$, and magnetisation-like $\mathcal{M} = \bar{\psi}\psi + s \cdot S$ operators, in analogy with the Ising
model and as applied to critical endpoints in QCD and the electroweak theory, \refer\cite{Rummukainen:1998as,Karsch:2001nf}, respectively. Matching the shape of the distribution to the universal one provides an additional handle on determining the location of the critical point and its scaling directions \cite{KAJANTIE1997413}. In particular, the universal scaling behaviour of energy-like observables depends crucially on approaching the critical point in the correct scaling direction.  

\begin{figure}[t]
    \centering
    \includegraphics[width=1 \textwidth]{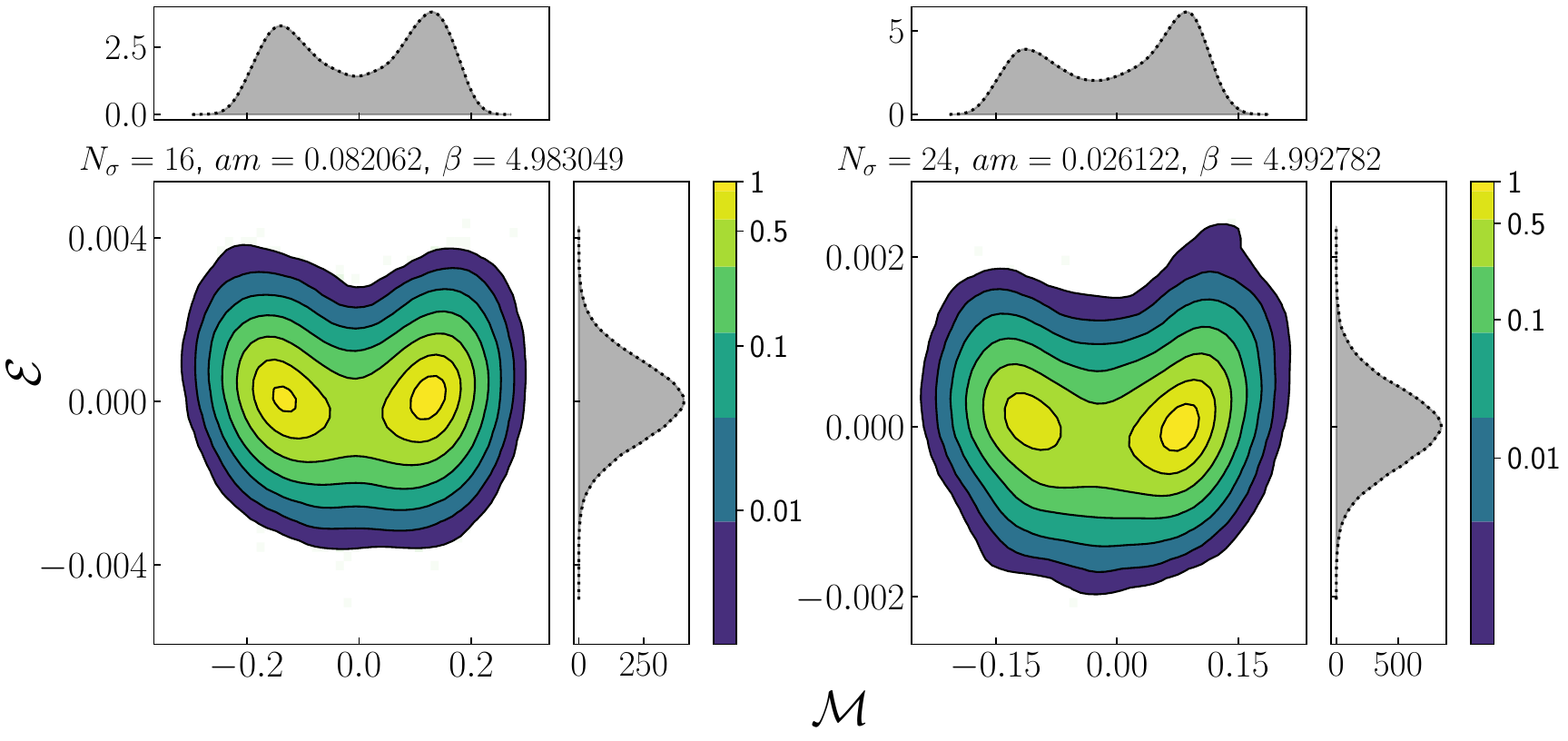}
    \caption{Joint distributions of the magnetisation-like $\mathcal{M}$ and energy-like $\mathcal{E}$ operators mapped onto their principal axes, as predicted by the MAF model at $(\amc,\betac) = (0.082062, 4.983049)$ for $\nt=4$ (left) and 
    $(0.026122, 4.992782)$ for $\nt=6$ (right). The characteristic shape of the 
    distribution in the principal-axis frame is universal at a $\ztwo$ 
    critical point, providing an independent handle on the location of the critical point and its scaling directions~\cite{Karsch:2001nf}.}
    \label{fig:outlook}
\end{figure}

Two examples of critical distributions extracted from our data for $\nt =4,6$ and mapped onto 
their respective principal axes, are shown in figure~\ref{fig:outlook}. 
We chose two specific ML runs from the test cases displayed as data points in 
figure~\ref{fig:amc-comparison} to plot their distributions at criticality
in figure~\ref{fig:outlook}.
The distribution plotted in the left panel of figure~\ref{fig:outlook} corresponds to 
the $\nt=4$ case testing $\ns$ interpolation using $8$ MADE blocks (grey points in figure~\ref{fig:amc-comparison} (left)).
The run with
the critical mass closest to the lattice reference value was selected.
The distribution plotted in the right panel of figure~\ref{fig:outlook} corresponds to
the $\nt=6$ case testing $\beta$ interpolation using ten MADE blocks with additional
data being generated at mass values that were not simulated (violet points in figure~\ref{fig:amc-comparison} (right)).
Again, the run whose critical mass is closest to the lattice reference value was
selected.
The location of the respective critical points is given by $(\amc,\betac)$,
whose exact values are given in the caption of figure~\ref{fig:outlook}.

\section{Conclusions}

In this work, we have explored machine-learning techniques to interpolate between distributions 
of lattice observables used for finite-size scaling analyses of the QCD phase structure, varying parameters such as the bare quark mass, gauge coupling, and spatial lattice volume. 
To this end, we have employed an implementation of conditional Masked Autoregressive Flows -- a framework well suited to probability density estimation from samples, conditioned on specified external parameters. 
Following up on first explorations~\cite{Neumann:2023,Karsch:2022yka}, we 
provide a systematic study of ML-based interpolation techniques used for the extraction of the $\ztwo$-critical mass separating first-order chiral phase transitions from crossovers. Specifically we considered unimproved staggered lattice QCD with five quark flavours, for which high-statistics data sets exist as 
benchmarks~\cite{Cuteri:2021ikv}.

In a first step, we successfully replace standard reweighting techniques in the lattice gauge coupling by ML-based interpolation. Second, we demonstrate by comparison with 
data subsets omitted from training, that this technique extends
to also interpolate in mass or spatial volume, for which reweighting methods
are computationally expensive or do not apply at all. Finally, combining 
interpolated and simulated data, estimates of the critical quark 
mass can be obtained with a reduced set of simulations.

However, we observe a systematic bias of ML-assisted 
critical masses towards smaller values, compared with those obtained exclusively
from simulation data. This bias can be understood by
the known mode-covering effect, which artificially
bridges bi-modal distributions, as occur in first-order transitions, 
and which we reproduce in Gaussian mixture models.
Although this mode-covering effect is characteristic of maximum likelihood training and does not vanish with extended training, we observe improved agreement with the lattice critical mass when increasing model complexity (MADE blocks).

At the present stage, this systematic error of the learned distributions precludes extensive use of the ML approach for precision measurements of critical endpoints
of first-order transitions.
However, the method is still useful in localising phase boundaries or criticality 
in terms of the relevant lattice parameters, thereby guiding subsequent high-precision measurements using MC simulations. It can also be used to identify 
the universal scaling axes in distributions interpolated to criticality.
Our study strongly motivates further work aimed at avoiding the 
mode-covering effect in bi-modal distributions. 

Additionally, note that the same model architecture trained on $\nt=4$ and $\nt=6$ data yields consistent results across both, as shown in figure~\ref{fig:amc-comparison}. This raises the question of whether critical quark mass predictions without this bias might also permit interpolation in $\nt$. If this were to be possible, it would allow to skip an entire sequence of volumes and associated masses, thus leading to substantial savings in computer time.

\acknowledgments
This work was supported by the Deutsche Forschungsgemeinschaft (DFG), German Research Foundation,  project numbers 315477589 (CRC-TR 211) and 460248186 (PUNCH4NFDI).
This project was also supported by the DFG as part of the CRC 1639 NuMeriQS -- project no.\ 511713970.
The authors thank the Bielefeld HPC.nrw team for their support and gratefully acknowledge access to the MLGPU Partition of the Marvin cluster of the University of Bonn, where most of the ML analysis was performed.
The code used for the machine learning analysis in this paper is adapted 
from the PhD project of Marius Neumann~\cite{Neumann:2023}.
Jan Philipp Klinger and Reinhold Kaiser acknowledge support by the Helmholtz Graduate School for Hadron and Ion Research (HGS-HIRe).

\appendix

\section{Neural network parameters}\label{sec:appendixNNparams}
In this section, we briefly discuss the choice of network parameters used in this analysis. Since this work derives from \cite{Neumann:2023,Karsch:2022yka}, most of the parameters, like choice of learning rate (LR) scheduler, optimiser, and values of L$1$, L$2$ regularisers, are kept unchanged. Some of the parameters, like the number of MADE blocks, their hidden units, batch size, training epochs, and range of epochs for the LR scheduler were tuned after performing validation tests on the lattice data and on the Gaussian mixture model (GMM) described in appendix~\ref{sec:AppB}. The summary of the parameters used is given in table~\ref{tab:maf_params}. 
\begin{table}[t]
\centering
\caption{Network and training parameters of the MAF model}
\label{tab:maf_params}
\begin{tabular}{ll}
\hline
\textbf{Network Parameter} & \textbf{Value} \\
\hline
Kernel regulariser & L$1$ \& L$2$ \\
L$1$ & $0.0001$ \\
L$2$ & $0.0001$ \\
Num MADE blocks & $8$, $9$, $10$ \\
Hidden units (per MADE) &  [128] \\
Activation & ReLU \\
Input dim & 2 \\
Conditional input dim & 3 \\
batch size & 2048 \\
Between-block transform & Reverse permutation \\
Loss & negative log likelihood \\
Optimiser & Adam \\
LR schedule & PolynomialDecay over 500 epochs, power=0.5 \\
Default LR endpoints & \texttt{base\_lr}=1e-3, \texttt{end\_lr}=1e-4 \\
Training Epochs & $500$ - $550$ \\
Samples for cumulants & $1$M \\
Samples for MMD & $100$k \\
\hline
\end{tabular}
\end{table}
\FloatBarrier

\begin{figure}
    \centering
    \includegraphics[width=0.9 \textwidth]{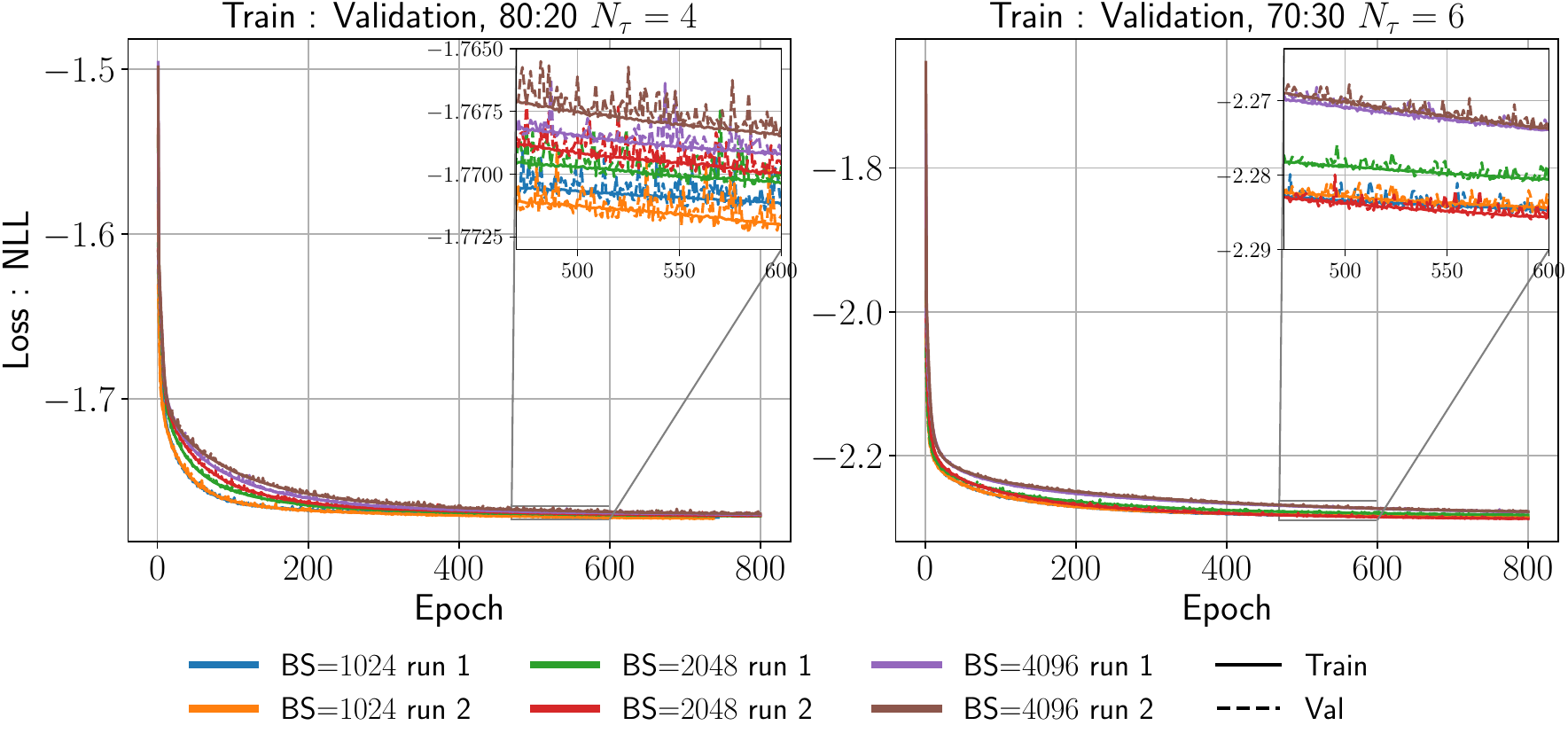}
    \caption{Loss convergence for (left) $\nt=4$ and (right) $\nt=6$. In both cases, we have split the data into train:validation sets and shown unbiased learning from different runs. We have further shown a comparison of different batch sizes.}
    \label{fig:LossCurves}
\end{figure}

\begin{figure}
    \centering
    \includegraphics[width=0.49 \textwidth]{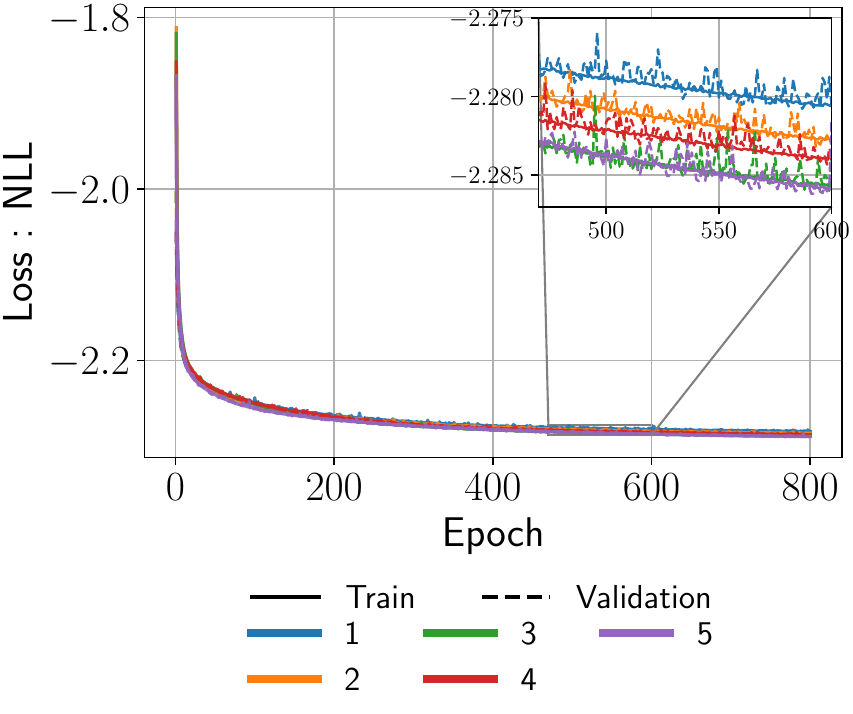}
    \caption{$\nt=6$ loss variation with runs.}
    \label{fig:LossNt6runs}
\end{figure}

To provide some justification for the training epochs, we show the training loss and validation in figure~\ref{fig:LossCurves}. For the $\nt=4$ data, the training set was split into $80 \%$ training and $20 \%$ validation sets (by assigning 20\% of each MC history to the validation set). The loss curves for two independent training runs are shown in the left plot of figure~\ref{fig:LossCurves} to show the lack of training bias. Based on the loss curves, it is hard to see any improvement after roughly $400$ epochs. However, when comparing the evaluated cumulants to the lattice data, we noticed we still needed to train up to $\sim500$ epochs. Values larger than $600$ were computationally expensive and also led to overfitting. Based on the data set, we chose to either train with $500$ or $525$ epochs. For the $\nt=6$ data, we split the data into $70 \%$ training and $30 \%$ validation sets. The loss curves in this case are shown in the right plot of figure~\ref{fig:LossCurves}, where we have also shown more runs to show the lack of bias between different training runs. We further compare different batch sizes to justify our choice of $2048$, which represents a compromise between reducing gradient noise through larger batches and controlling the associated training and memory costs. Using figure~\ref{fig:LossNt6runs}, we illustrate the convergence of different models toward a common minimum for a fixed batch size. 

\subsection{Technical details}
For the purpose of the ML analysis, we have exclusively made use of the TensorFlow library \cite{tensorflow2015-whitepaper}. Additionally, the entire ML workflow -- including training and sampling was performed on the MLGPU partition consisting of A40 GPUs, on the Marvin HPC cluster at the University of Bonn.

A useful hardware-specific metric to describe the training time on the A40 GPUs for this algorithm would be the training time per step. Using this, one can construct the time for the entire training process using the formula
\begin{equation}
    t_{\mathrm{full \, training}} = t_{\mathrm{per \, step}} \times \frac{N_{\mathrm{data}}}{N_{\mathrm{batch}}} \times N_{\mathrm{epochs}} \,.
\end{equation}
Training times per step were typically between $11$ and $13$ ms. The total data used for each test, (i) $\beta-$interpolation, (ii) $am$-interpolation, and (iii) $\ns$-interpolation can be found in table~\ref{tab:nt4-statistics}. A final ingredient of this computation is the batch size, which we fixed to $N_{\mathrm{batch}} =2048$. This gave us a total training time of roughly $4-6$ hours. 
Sampling is very cheap, and for each parameter set, $\beta$, $am$, and $\ns$, it takes roughly $5$ s to sample $1$M points. For the analysis described above, this implies a sampling time of roughly $13$ minutes if samples are generated at $200$ $\beta$ values at each of the four mass values and at each of the three volumes.

\section{Tests on Gaussian mixture models}\label{sec:AppB}
In this section, we would like to address some general questions on the performance of the MAF network used in this work, including the choice of network parameters. For this, we need a distribution of the form $p(x_1,x_2 \mid y_1, y_2, y_3)$, which has two input dimensions while depending on three conditional inputs. To have a closer resemblance to lattice data describing a change from a crossover to a first-order transition, we would additionally like our distribution to have attributes like mode separation and correlations between inputs, while being able to interpolate meaningfully in the conditional variables. A natural choice for such a distribution is the two-component Gaussian mixture model (GMM)~\cite{bishop2006pattern}, parametrised by the mean parameter $a$, width parameter $b$, and the correlation between the two dimensions, given by $c$. 

The two-component GMM is described by the probability density
\begin{equation}
    p(x_1,x_2 \mid a, b, c) = s \cdot  \mathcal{N}(\bm{x} \mid \bm{\mu}_1, \mathbf{\Sigma}) + (1-s) \cdot \mathcal{N}(\bm{x} \mid \bm{\mu}_2, \mathbf{\Sigma})\;,
\end{equation}
 with input $\bm{x}  = [x_1,x_2]^T$, means $\bm{\mu}_1 = [-a, 0]^T \,,\,\bm{\mu}_2 = [a, 0]^T$, and identical covariance matrix for both,
\begin{equation*}
\mathbf{\Sigma} = \begin{bmatrix}
b^2 & c \\
c & b^2
\end{bmatrix}.
\end{equation*}
The variable $c$ measures the correlation between the two input dimensions $x_1,x_2$, with $c=0$ giving uncorrelated samples along each dimension. The mixing coefficient between the two Gaussians is set to $s = 0.5$ and is implemented using a Bernoulli trial. For each assigned component, a sample is drawn from the corresponding multivariate normal distribution with mean $\bm{\mu}_i$ and covariance $\mathbf{\Sigma}$.

 In order to test our network on this distribution, we construct mixtures with the parameters listed in table~\ref{tab:param_list}, chosen to satisfy the positivity constraint for $\Sigma$ using $| c | < b^2$. The parameters are chosen to represent a transition from an easy distribution representing two overlapping Gaussians to widely separated Gaussians with high correlation between the input dimensions. Furthermore, three test cases are considered, with $10$k, $20$k, and $40$k samples each, for all the parameter sets shown in table~\ref{tab:param_list}. Different models were trained on \emph{(1)} each of the three sample sizes and \emph{(2)} using different numbers of MADE blocks and widths of each layer. In all cases, the network was trained simultaneously on the GMMs represented by the parameters in table~\ref{tab:param_list}.

\begin{table}[h]
\centering
\begin{tabular}{ccc || ccc}
\hline
$a$ & $b$ & $c$ & $a$ & $b$ & $c$ \\
\hline
0.2 & 0.5 & 0.0  & 1.0 & 0.8 & 0.2 \\
0.2 & 0.5 & 0.2 & 1.5 & 0.6 & -0.3 \\
0.6 & 0.8 & 0.4   \\
\hline
\end{tabular}
\caption{Parameter sets $(a,b,c)$ used in the study of the GMM (split over two columns).
Three test cases are considered, with $10$k, $20$k, and $40$k samples for each parameter set, respectively.}
\label{tab:param_list}
\end{table}

\subsection{Results: learning the distribution} \label{sec:GMM_dist}
\begin{figure}[ht]
    \centering
    \includegraphics[width=1 \textwidth]{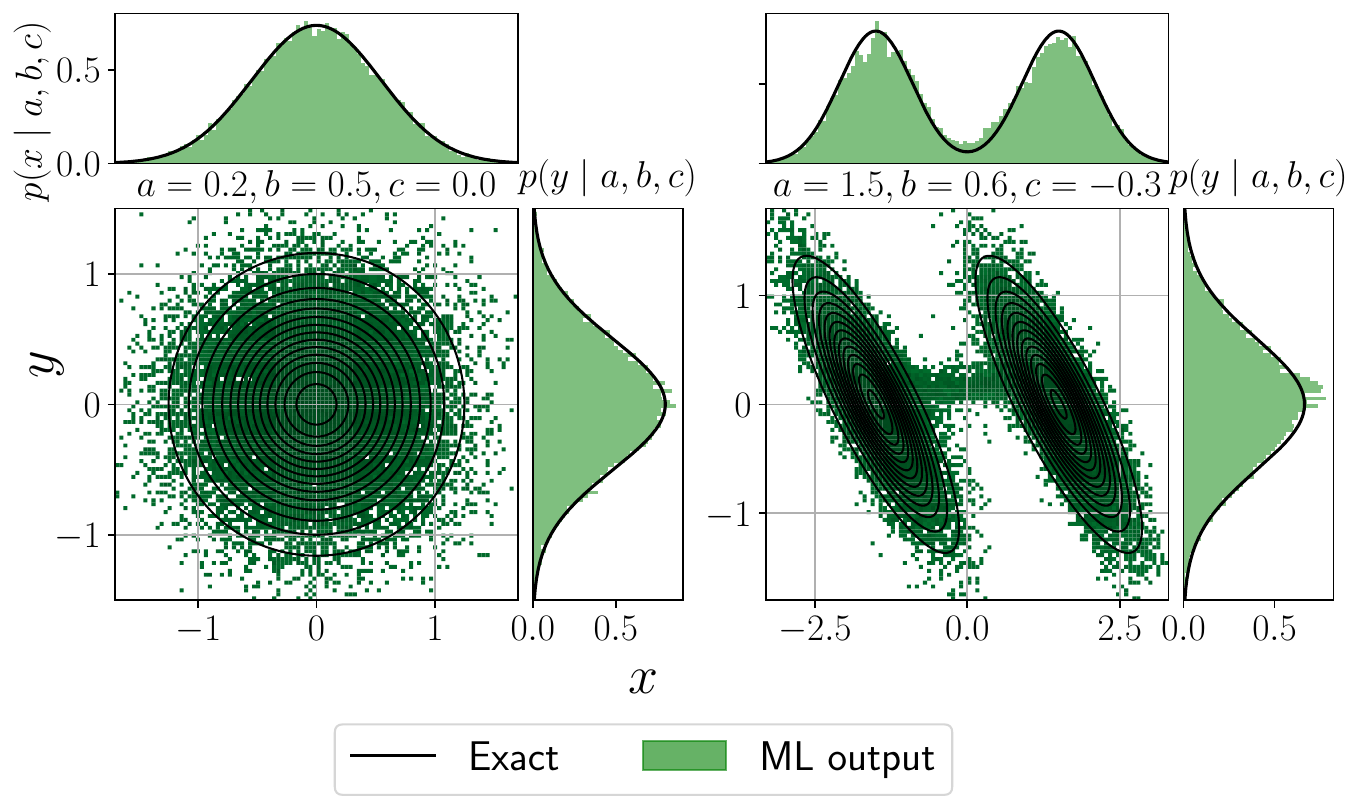}
    \caption{Comparison of samples from the MAF model trained with size $20$k, ten MADE blocks, and width $N=128$ in the hidden layer of each block (see figure~\ref{fig:MAF}) against contours of the exact distributions: (left) the ``easiest'' distribution with $(a,b,c)=(0.2,0.5,0.0)$ and (right) the ``hardest'' distribution with $(a,b,c)=(0.6,0.8,-0.3)$, along with the marginals.}
    \label{fig:GMM_20k_densitycomp}
\end{figure}

In this section, we show the results of the training procedure for one choice of the various cases listed above, particularly the case with ten MADE blocks, sample size $20$k, and width of the network $N=128$. We emphasise that results from all other cases are visually indistinguishable from this choice shown in figure~\ref{fig:GMM_20k_densitycomp} \footnote{Data for plots available on GitHub~\cite{singh2026lqcdmaf}}. In the figure, we point to one feature known in the community as ``mode covering'' \cite{Minka2005Divergence,Hackett:2021idh,Nicoli:2023qsl}. This occurs when a generative model is trained by maximising the likelihood of the data. The model is strongly penalised if it assigns very small probability to configurations that appear in the ensemble. As a result, the model tends to distribute probability mass so as to cover all regions where data are present. In practice, this often leads to slightly broadened distributions: rather than missing a peak, the model smooths across neighbouring regions and may overestimate fluctuations between modes. This behaviour is evident in the right plot of figure~\ref{fig:GMM_20k_densitycomp}, where the model successfully recovers the two separated modes but introduces an artificial bridge of probability mass between them.

\subsection{Comparing learning across parameters: maximum mean discrepancy}\label{sec:GMM_MMD}
The goal of this section is to quantify the relative learning between the different cases mentioned above.  For this, we need to choose a metric that can quantify how different two given distributions are. In the literature, one usually uses the Kullback--Leibler divergence
\cite{Kullback1951} to measure the distance between two distributions, however for this one needs to know the normalisations of the distributions. 

\begin{figure}[ht]
    \centering
    \includegraphics[width=1 \textwidth]{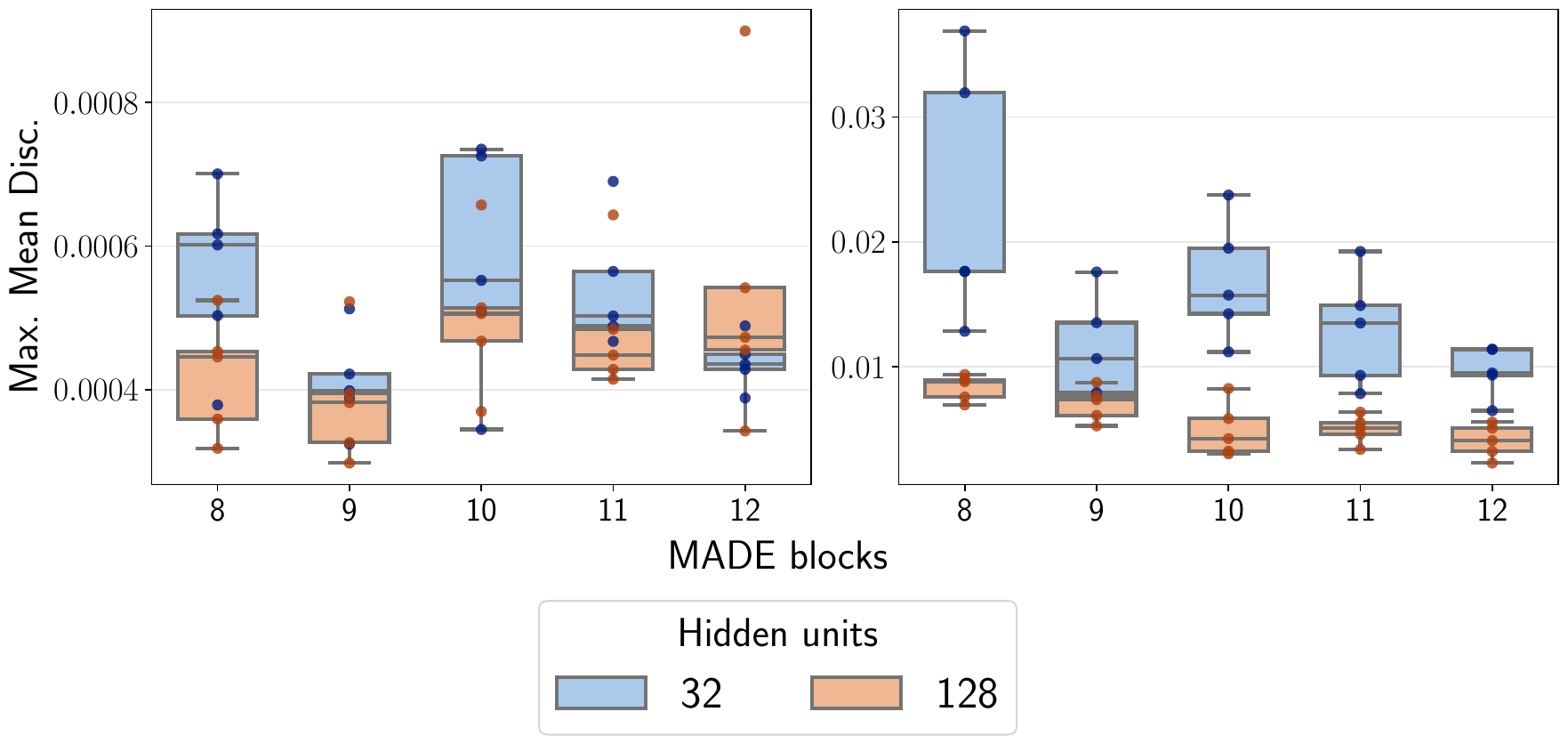}
    \caption{ Comparison of MMD with the easiest (left) to most difficult (right) distribution with MAF trained on $20$k samples with MADE layer width $N=32$ (blue) and width $N=128$ (orange), plotted against the number of MADE blocks.}
    \label{fig:GMM_made20k_mmd}
\end{figure}

\begin{figure}[ht]
    \centering
    \includegraphics[width=1 \textwidth]{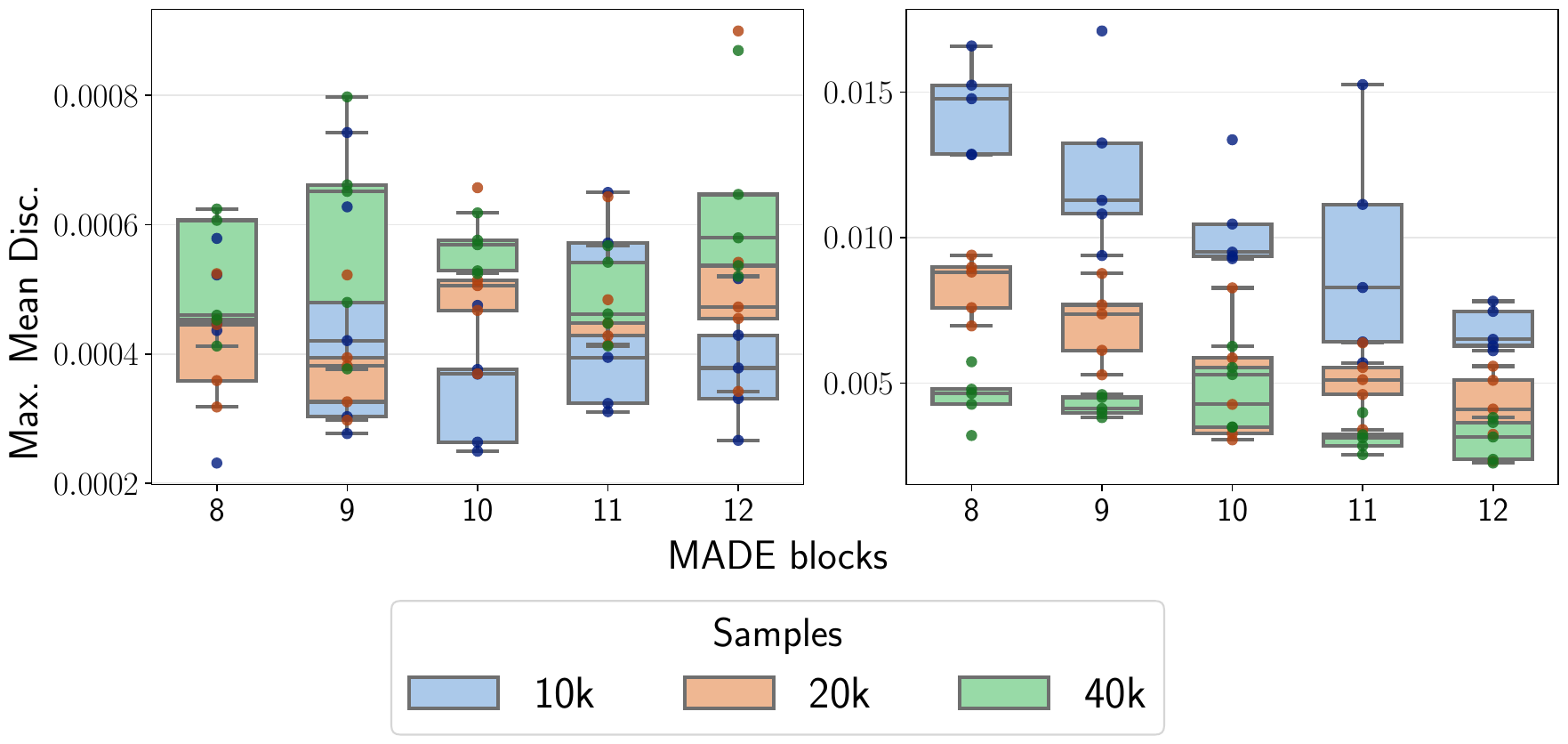}
    \caption{Comparison of MMD with the easiest (left) to most difficult (right) distribution with MAF trained on $10$k, $20$k, and $40$k samples with fixed width $N=128$, plotted against the number of MADE blocks.}
    \label{fig:GMM_made_samples_mmd}
\end{figure}

In this work, we choose to compute the maximum mean discrepancy (MMD) \cite{JMLR:v13:gretton12a}, as a nonparametric two-sample test to quantitatively compare the true data (lattice simulations) with the samples generated by the machine learning model. The MMD provides a scalar measure of discrepancy between the two empirical distributions and thus allows us to assess whether the two sample sets are statistically consistent with having been drawn from the same underlying probability distribution. We have adapted the implementation of the MMD algorithm as given in \cite{tunali2019mmd}. In Figs.~\ref{fig:GMM_made20k_mmd} and \ref{fig:GMM_made_samples_mmd}, we show the relative training differences between the various parameter sets of the GMM model. In each of these plots, the distributions of the MMD values obtained from multiple independent runs are visualised using box plots produced with the \texttt{seaborn} visualisation library~\cite{Waskom2021}. Each box indicates the interquartile range (IQR) between the first and third quartiles, while the horizontal line denotes the median. The whiskers extend to the most extreme values within $1.5\times$ IQR, and points beyond this range are shown as outliers. Individual run results are additionally overlaid as markers to illustrate the full distribution of the data.

In figure~\ref{fig:GMM_made20k_mmd}, we show how the relative training between the parameters of the GMM depends on the choice of network parameters like the number of MADE blocks and the number of hidden units in each MADE block. For this, we remind the reader that one model is trained on \emph{all} parameter values shown in table~\ref{tab:param_list}, for each choice of MADE blocks and hidden units shown. As increasing these parameters increases the complexity of the model, our goal is to determine how model complexity affects training. Looking at the left plot in figure~\ref{fig:GMM_made20k_mmd}, we notice that for this parameter choice $(a=0.2,b=0.5,c=0)$ there is no advantage in either increasing the number of hidden units or MADE blocks. However, looking at the right plot of the figure, with parameter choice $(a=1.5,b=0.6,c=-0.3)$, we see that increasing the model complexity is clearly advantageous. 

In figure~\ref{fig:GMM_made_samples_mmd}, we perform a test to study the effect of increasing the number of samples with a fixed number of hidden units but increasing the model complexity in terms of the number of MADE blocks. Again, we see that for the ``easiest'' distribution with $(a=0.2,b=0.5,c=0)$, increasing the number of input samples or the number of MADE blocks has no clear effect on the learning. On the other hand, in the right plot, we see a clear advantage both when increasing the number of input samples and the number of MADE blocks.
\FloatBarrier

\section{Extended results on learned distributions for \texorpdfstring{\boldmath$\nt=4$}{Nt=4}}\label{sec:appFigs}
In this section, we provide additional figures for the cumulants for $\nt = 4$, along with the corresponding results of the MMD analysis (described in section~\ref{sec:GMM_MMD}), to help the reader interpret the results shown in figure~\ref{fig:amc-comparison}. This section contains three figures, each showing in the top panel the comparison of the ML-learned distribution against the MC data using the MMD metric and, in the bottom panel, the cumulant analysis for all the masses and volumes present in the $\nt=4$ data set. Each column of the figure represents results for a single spatial lattice, with $\ns=8$ on the left, $\ns=12$ in the center, and $\ns=16$ on the right and contains masses coloured blue, orange, green, and red for $am = \{0.075,0.080,0.085,0.090 \}$, respectively. Every row corresponds to a different cumulant, with the mean of the chiral condensate in the first row, followed by the susceptibility, skewness and kurtosis in the following rows.

\begin{figure}
    \centering
    \hspace{-2mm}\includegraphics[width= 1.008\textwidth]{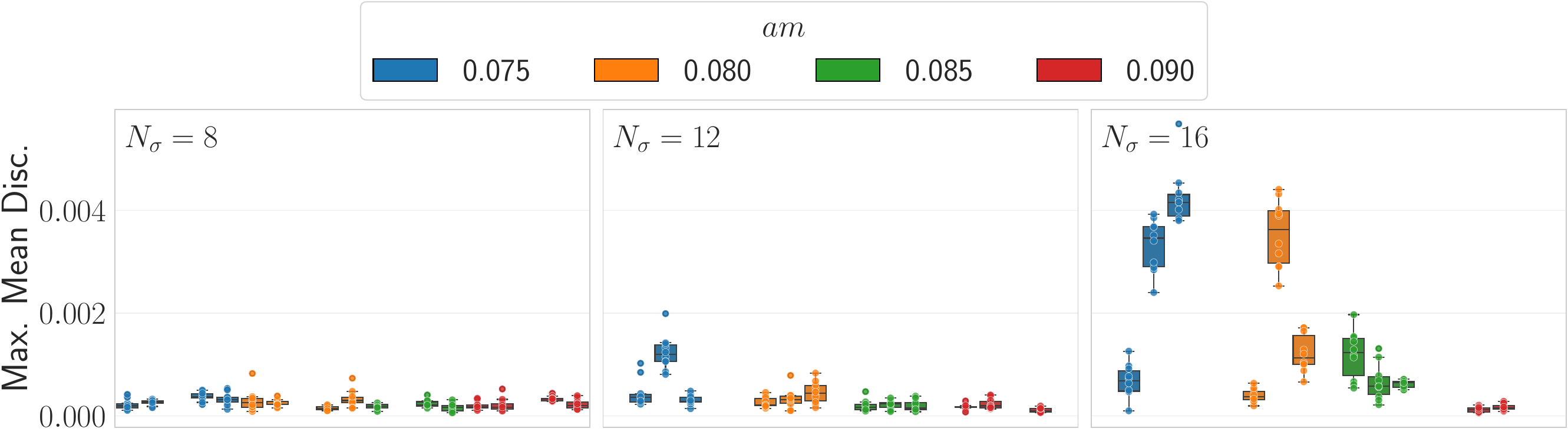}
    \includegraphics[width= 1\textwidth]{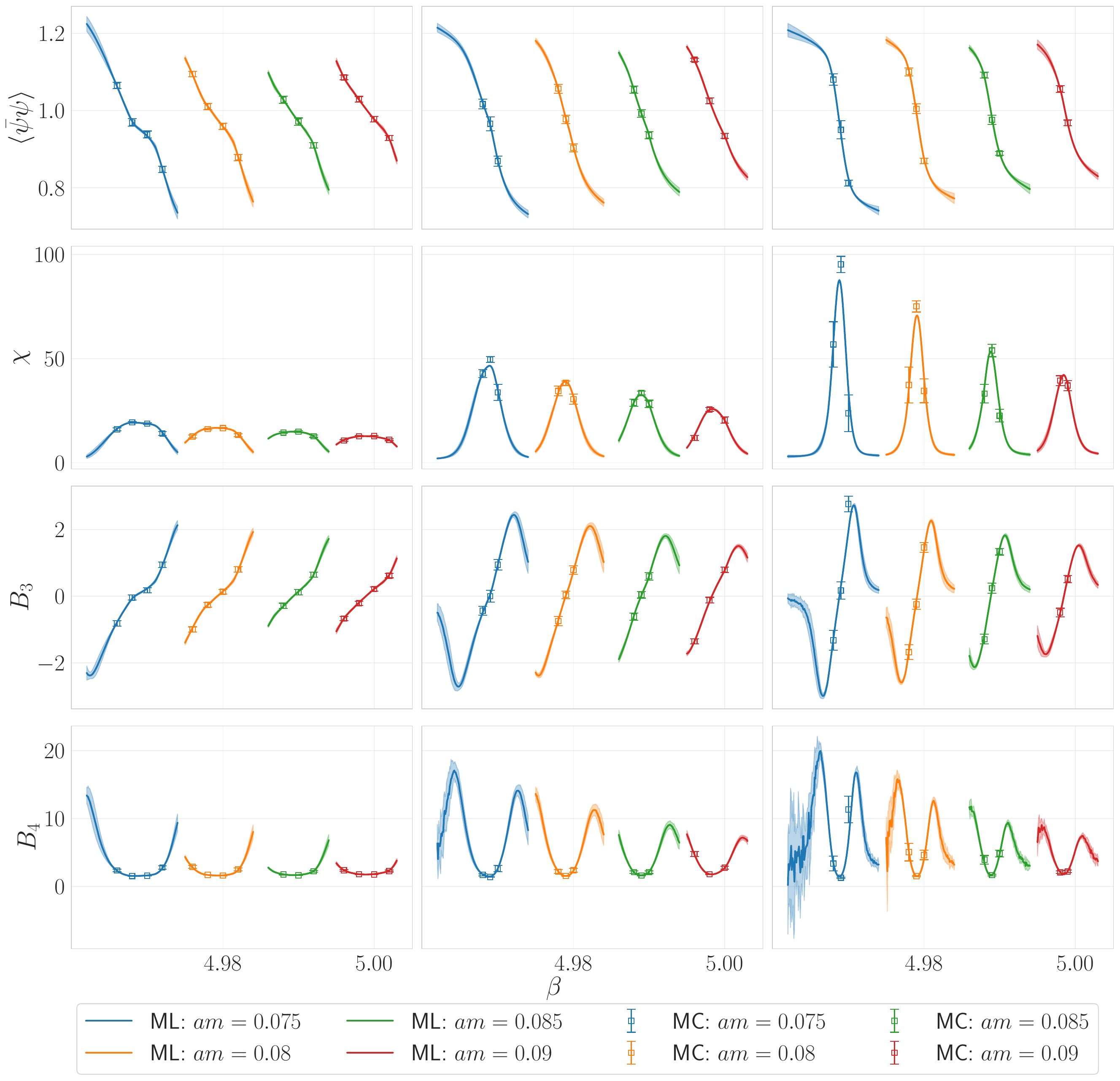}
     \caption{Interpolation in coupling $\beta$ (corresponding to figure~\ref{fig:beta_interpolation}). From left to right, we show the cumulants computed from the MAF training for $\ns \in \{8,12,16 \}$. In the top panel, we show the related MMD distribution for each volume. The model was trained on all data for every volume and mass.}
    \label{fig:cumulantNt4_all}
\end{figure}

\begin{figure}
    \centering
    \includegraphics[width= 1\textwidth]{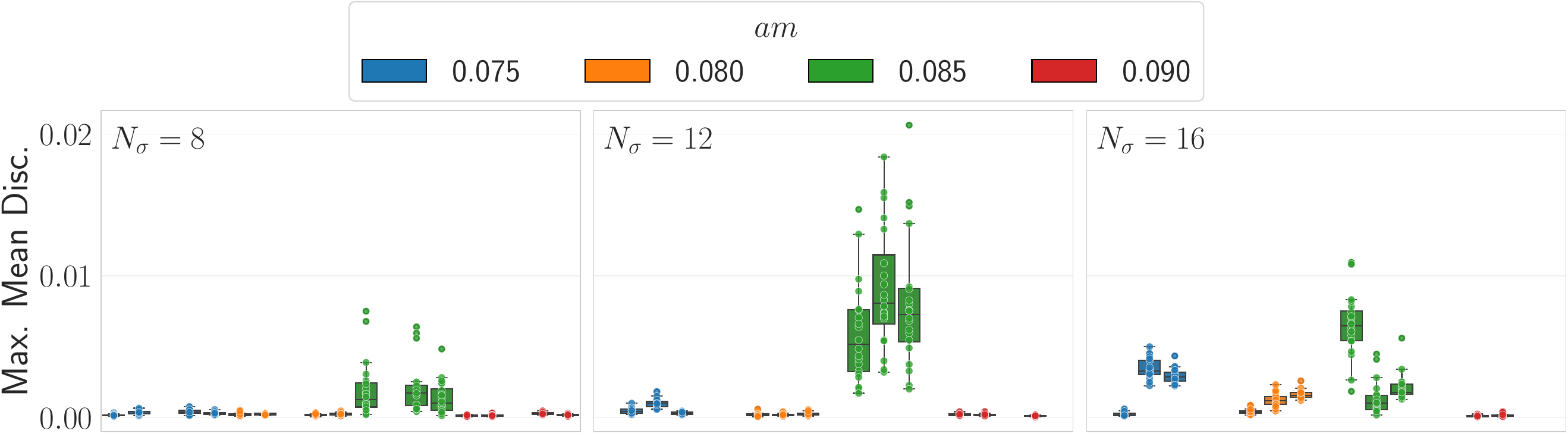}
    \includegraphics[width= 1\textwidth]{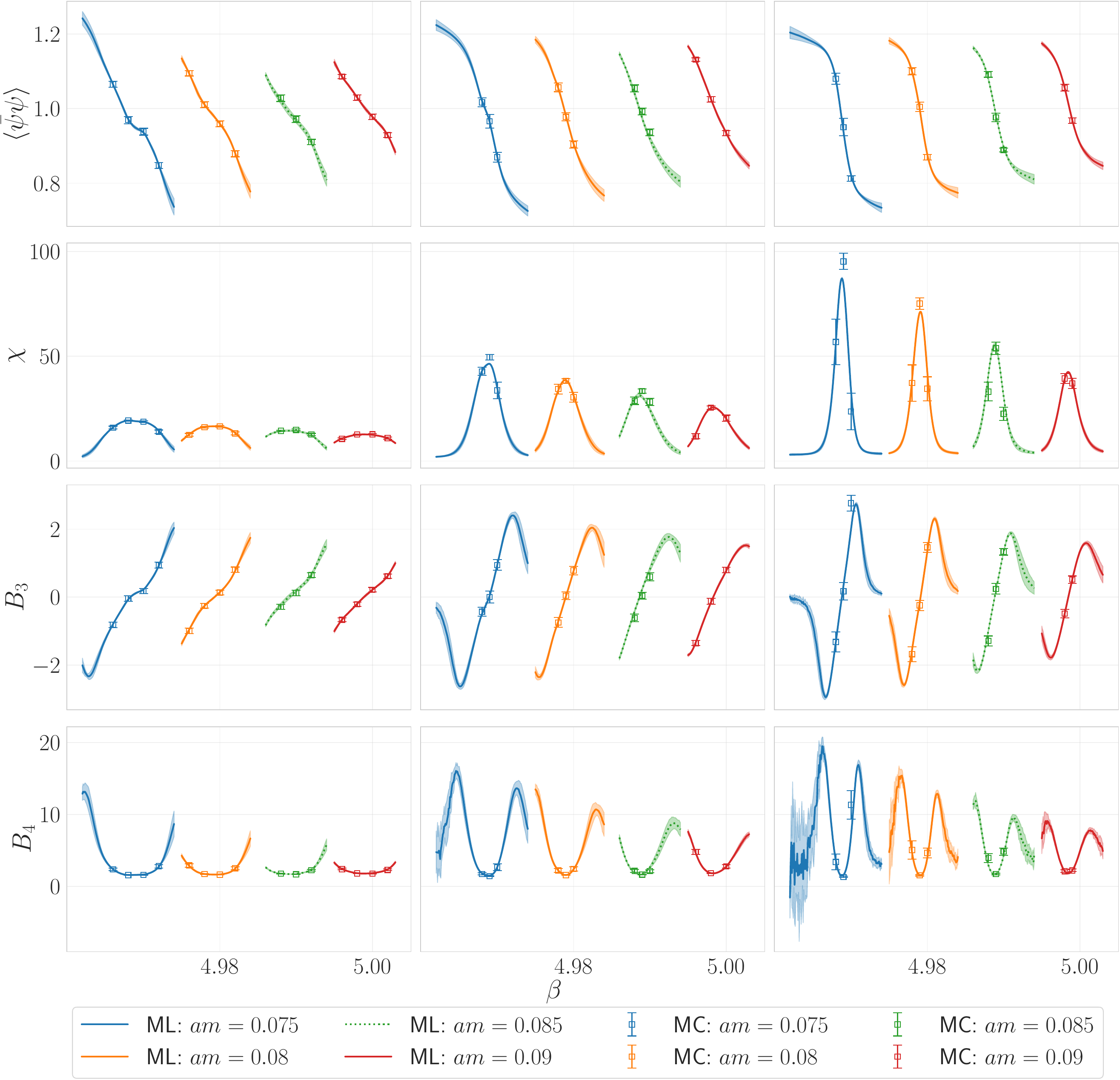}
     \caption{Interpolation in mass $am$ (corresponding to figure~\ref{fig:mass_interpolation}). From left to right, we show the cumulants computed from the MAF training for $\ns \in \{8,12,16 \}$. In the top panel, we show the related MMD distribution for each volume. The model was trained on all data except $am=0.085$ (dotted). }
    \label{fig:cumulantNt4_remM085}
\end{figure}

\begin{figure}
    \centering
    \includegraphics[width= 1\textwidth]{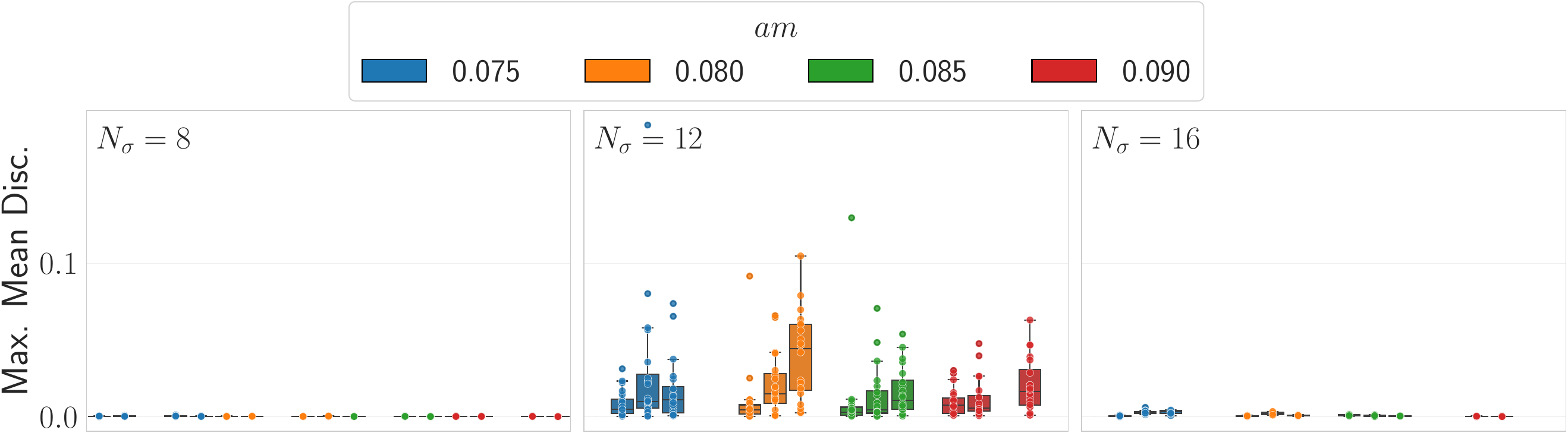}
    \includegraphics[width= 1\textwidth]{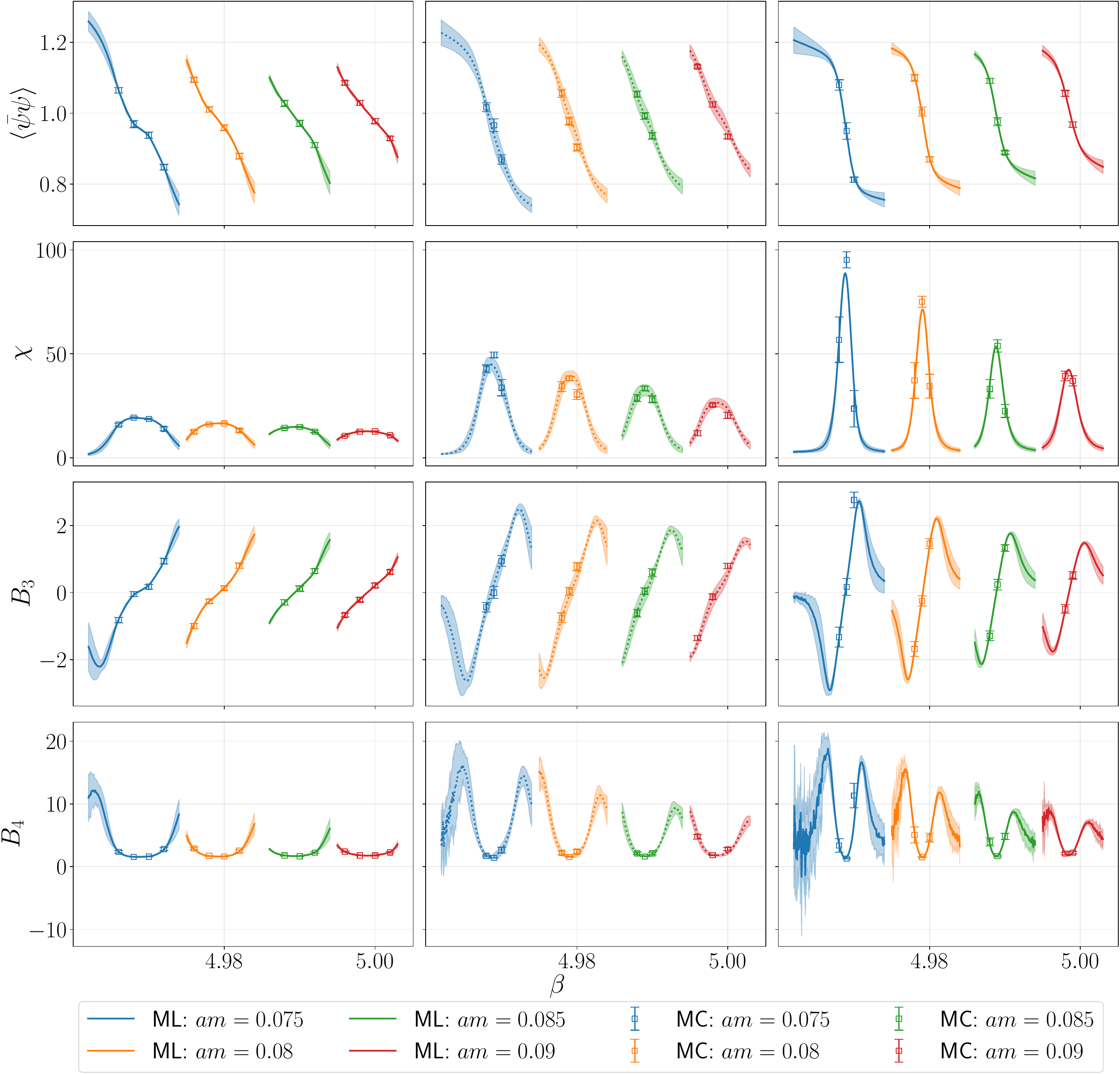}
     \caption{Interpolation in volume $\ns$ (corresponding to figure~\ref{fig:volume_interpolation_cumulants}). From left to right, we show the cumulants computed from the MAF training for $\ns \in \{8,12,16 \}$. In the top panel, we show the related MMD distribution for each volume. The model was trained on all data except $\ns=12$ (dotted).}
    \label{fig:cumulantNt4_rem12}
\end{figure}

In the top panel of figure~\ref{fig:cumulantNt4_all}, the deviations at the level of the distributions are shown by the MMD between the learned ML densities and all available MC distributions. We observe some trends, like an increase in the MMD values for $\ns=16$, indicating poorer learning of those distributions compared to the smaller volumes. This is expected, as the double-peak structure becomes more prominent for larger $\ns$ in the first-order region. We further observe that at large masses, where the transition is a crossover, the distributions are learned more accurately -- indicated by lower MMD values. In contrast, at smaller masses, the presence of a first-order transition, characterised by a pronounced double-peak structure, gives rise to more complex distributions that are harder for the MAF model to learn accurately. The panels showing the comparison at the level of averaged cumulants show good agreement with the lattice data. \\
In figure~\ref{fig:cumulantNt4_remM085}, we show the corresponding results for the case where the quark mass $am=0.085$ was removed across all $\ns$. However, since we have the MC data corresponding to that mass, we were also able to perform the MMD analysis for this case. From the top panel of figure~\ref{fig:cumulantNt4_remM085}, we observe that the general agreement of all learned distributions is worse compared to the case where all data was used during training (see the top panel in figure~\ref{fig:cumulantNt4_all}). Apart from this, the general trend of poorer learning for the largest $\ns=16$ still remains. The newest feature is that the MMD for the distributions corresponding to the mass values that were systematically removed across all $\ns$ is the highest -- as expected -- indicating the greatest disagreement in those distributions. The panels showing the comparison at the level of averaged cumulants show good agreement with the lattice data, with larger uncertainty for the omitted mass. \\
In figure~\ref{fig:cumulantNt4_rem12}, we show the corresponding results for the case where the data corresponding to $\ns=12$ was removed entirely. However, since the MC data corresponding to that mass is available, we were also able to perform the MMD analysis for this case. From the top panel of figure~\ref{fig:cumulantNt4_rem12}, we observe that the general agreement of all the learned distributions is worse compared to the case where all data was used (see the top panel in figure~\ref{fig:cumulantNt4_all}). Apart from this, the general trend of poorer learning for the largest $\ns=16$ still remains (now one should only compare $\ns=8$ with $\ns=16$). The newest feature is that the MMD for the distributions corresponding to $\ns=12$ is the highest -- which is expected as the data was not trained on it. The panels showing the comparison at the level of averaged cumulants show good agreement with the lattice data, with larger uncertainty for the omitted $\ns=12$ data sets.

\section{Complementary results for \texorpdfstring{\boldmath$\nt=6$}{Nt=6}}\label{sec:complementary_nt6}
In this section, we present results for the $\nt=6$ data that complement those for $\nt=4$, specifically the data corresponding to figures~\ref{fig:runs_sys_unv1},~\ref{fig:beta_interpolation}, and~\ref{fig:volume_interpolation_cumulants}.
These figures allow us to assess the quality of the data used to produce the $\nt=6$ results presented in section~\ref{sec:z2-results}.
So far no ML runs at $\nt=6$ exist for mass interpolation, which is the reason why the figure complementary to figure~\ref{fig:mass_interpolation} is omitted here.

\subsection{Multi-seed analysis for \texorpdfstring{$\nt=6$}{Nt=6} data}
\begin{figure}[t]
    \centering
        \includegraphics[width=1\linewidth]{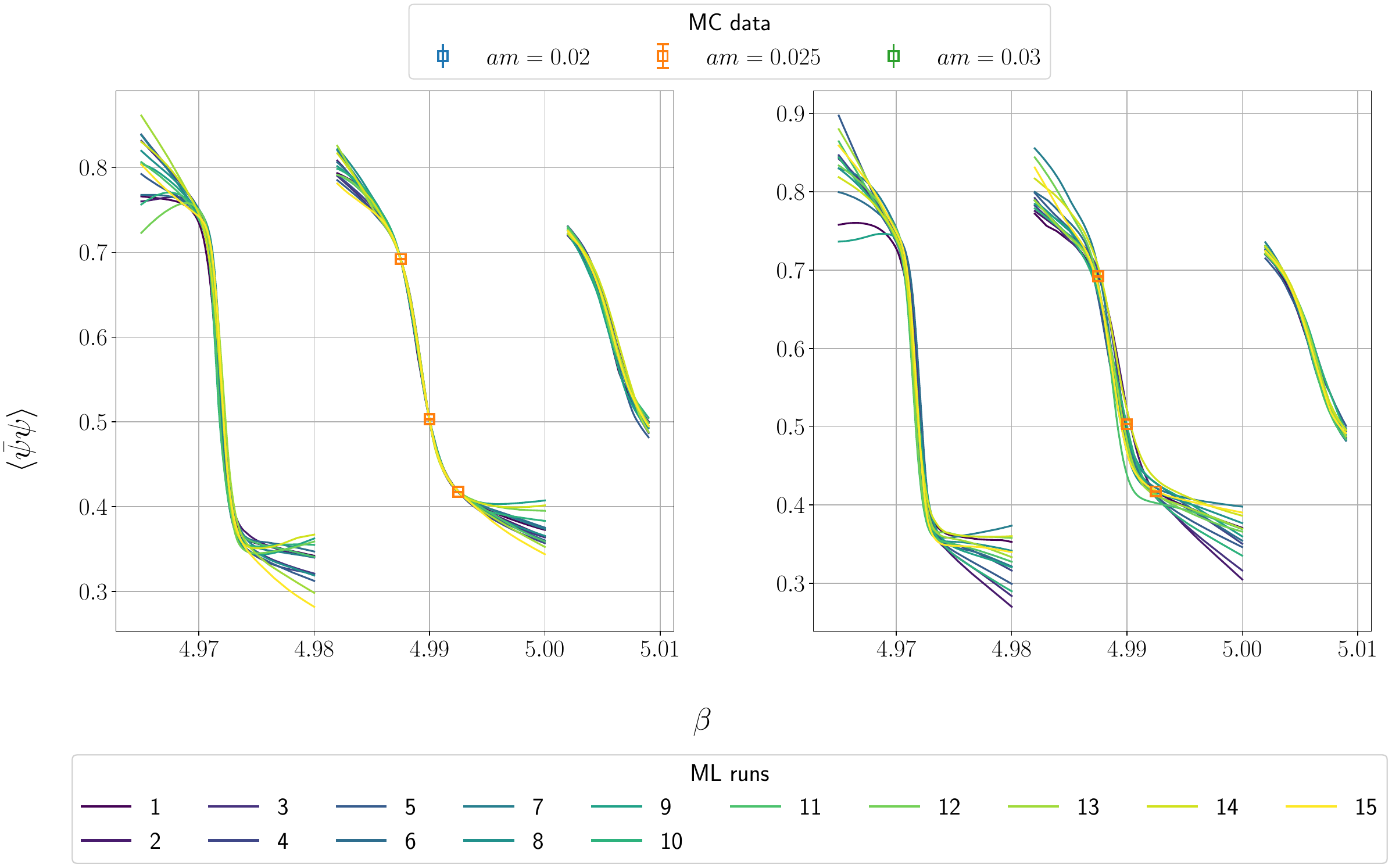}
    \caption{ML predictions across independent training runs compared with MC data for $\nt=6$, $\ns=16$ data. Left: training on all data yields consistent results across runs. Right: excluding the $\ns=16$ data leads to increased run-to-run variation, reflecting reduced constraints on the learned distribution while maintaining consistency with the MC data. Note that in contrast to $\nt=4$, we don't have MC data for all the masses $am \in \{0.02,0.025,0.03 \}$ for the volumes $\ns=16$ and $\ns=18$.}
    \label{fig:Nt6runTorunvar}
\end{figure}

In the spirit of sections~\ref{sec:sampling_error_one_model} and \ref{sec:variation_of_model_ensemble}, we estimate the seed-to-seed variability of the ML model against MC data for two cases: (i) training performed on the full data set, and (ii) training performed with an entire volume, here $\ns=16$, excluded. As shown in figure~\ref{fig:Nt6runTorunvar}, the results are qualitatively the same as those presented in figure~\ref{fig:runs_sys_unv1} for the $\nt=4$ case. The main difference is the smaller size of the data basis in this case, since fewer MC simulations were performed for the finer lattices $\nt=6$ (see table~\ref{tab:Nf5_statistics}).

\subsection{Interpolation in coupling}
\begin{figure}[t!]
    \centering

    \begin{subfigure}[t]{\textwidth}
        \centering
        \includegraphics[width=\textwidth]{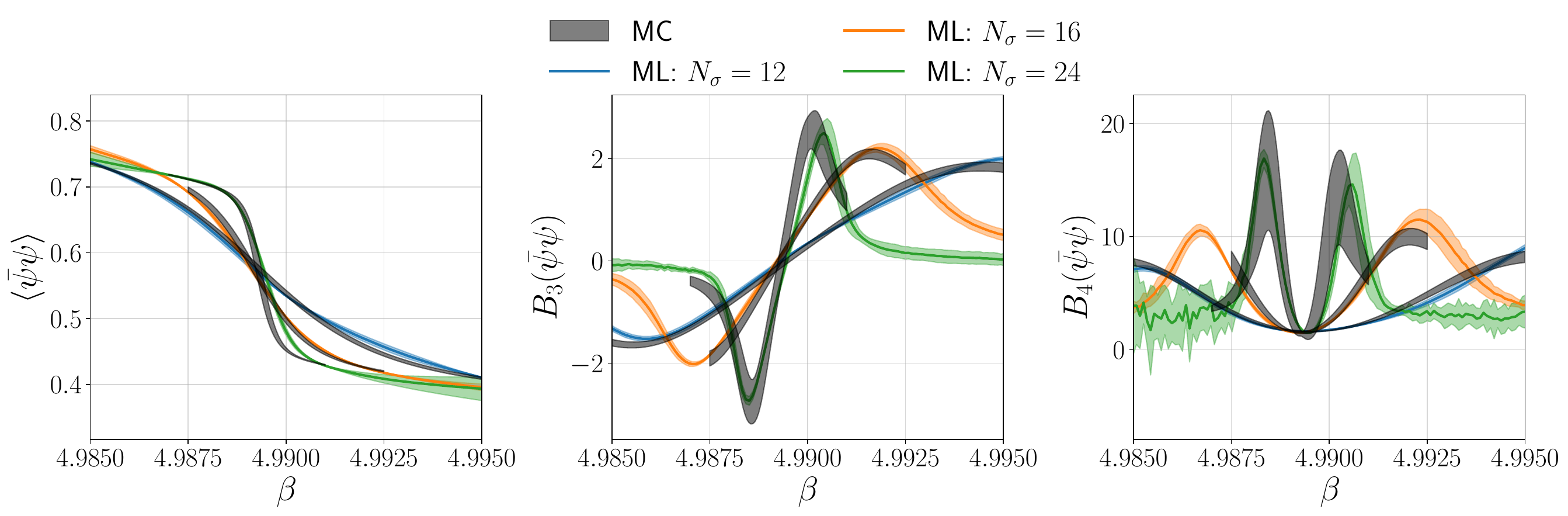}
        \caption{$am=0.025$}
        \label{fig:beta_interpolation_mass0025_nt6}
    \end{subfigure}

    \vspace{0.5cm}

    \begin{subfigure}[t]{\textwidth}
        \centering
        \includegraphics[width=\textwidth]{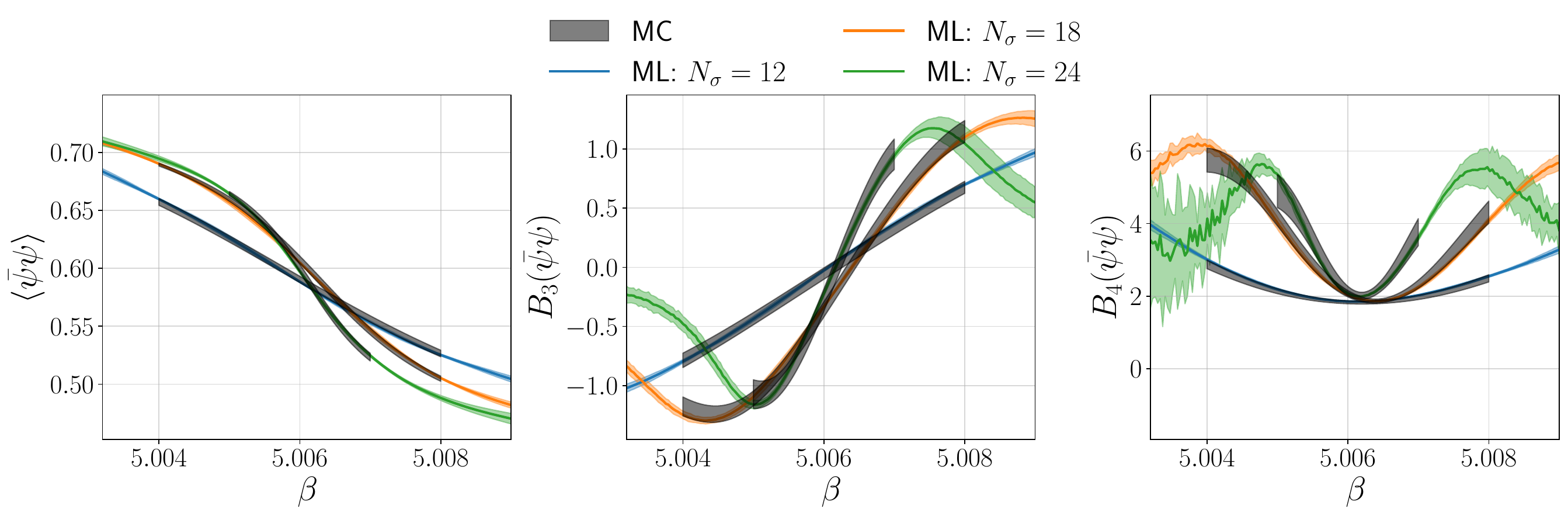}
        \caption{$am=0.030$}
        \label{fig:beta_interpolation_mass003_nt6}
    \end{subfigure}

    \caption{
        Interpolation in $\beta$ at $N_\tau=6$ for two different fermion
        masses. The mean $\langle\pbp\rangle$, skewness
        $B_3\left(\pbp\right)$, and kurtosis
        $B_4\left(\pbp\right)$ (\eq\eqref{equ:std-moments}) of standard
        reweighted MC data (black) are compared against ML-generated data
        for three different volumes.
        }
    \label{fig:beta_interpolation_nt6}
\end{figure}

Figure~\ref{fig:beta_interpolation_nt6} shows the interpolation in $\beta$ of the mean value $\ev{\pbp}$, the skewness $B_3(\pbp)$ and the kurtosis $B_4(\pbp)$of the chiral condensate at $\nt=6$ for two fermions masses.
The predictions from ML models trained on the full data set of $\nt=6$ are compared
with the results from standard multiple-histogram reweighting.
Like for the $\nt=4$ case in figure~\ref{fig:beta_interpolation}, good agreement is achieved near the transition point.

\subsection{Interpolation in volume}\label{sec:vinter_appendix}
\begin{figure}[t!]
    \centering
    \includegraphics[width=\textwidth]{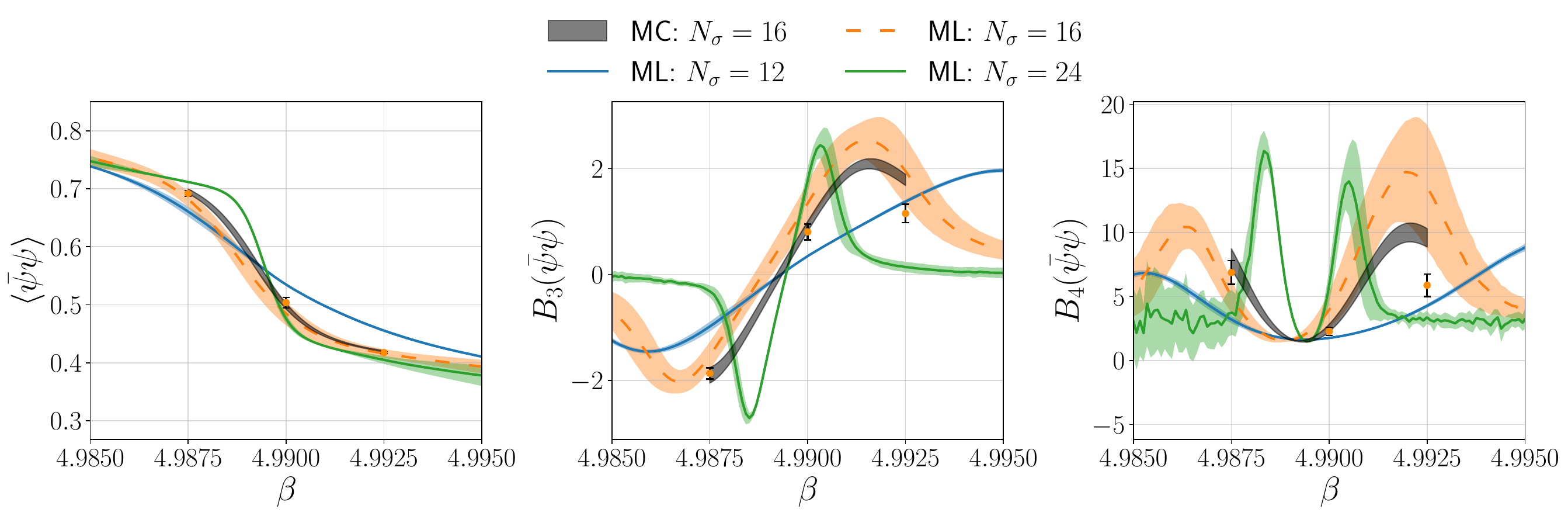}
    \caption{Interpolation in volume for $am=0.025$ and $\nt=6$. The mean $\langle  \pbp\rangle$, skewness $B_3\left(\pbp \right)$ and kurtosis $B_4\left(\pbp \right)$ (\eq\eqref{equ:std-moments}) of standard reweighted MC data (black) are compared with ML-generated data for three different volumes. Solid lines show the volumes $\ns=\{12,24\}$ which were included in the training data, the dashed line shows results for the volume $\ns =16$, which are obtained through interpolation.}
\label{fig:volume_interpolation_cumulants_nt6}
\end{figure}

Figure~\ref{fig:volume_interpolation_cumulants_nt6} shows the interpolation in $\ns$ of the mean value $\ev{\pbp}$, the skewness $B_3(\pbp)$ and the kurtosis $B_4(\pbp)$of the chiral condensate at $\nt=6$ for one fermion mass.
The predictions from ML models trained on the data set of $\nt=6$, with the entire volume $\ns=16$ excluded, are compared
with the results from standard multiple-histogram reweighting at $\ns=16$.
Except for a slight shift of the ML prediction towards smaller $\beta$ values, reasonable agreement is achieved near the transition point similar to the $\nt=4$ case shown in figure~\ref{fig:volume_interpolation_cumulants}.
An increased systematic deviation from the lattice results is expected, given the exclusion of $\ns=16$ from the data basis combined with the fact that the overall data basis for $\nt=6$ is already smaller than that for the $\nt=4$ case.
Note that the kurtosis finite-size scaling analysis, whose results are presented in section~\ref{sec:estimating-ztwo}, is not very sensitive to this shift, since it does not significantly affect the kurtosis value at the transition point, $B_4(\betapc)$.

\FloatBarrier
\bibliographystyle{JHEP}
\bibliography{biblio.bib}

\end{document}